\documentclass[preprint,12pt]{elsarticle}

\usepackage{amssymb}
\usepackage{amsmath}
\usepackage{cases}
\usepackage{bm}
\usepackage{color}

\newcommand\WR{{\rm W\hspace{-0.7mm}R}}
\newcommand\IPF{{\rm IPF}}
\newcommand\DR{{\rm DR}}

\journal{Journal of Computational Physics}

\begin{document}

\begin{frontmatter}



\title{{Boundary-velocity error} and stability of the accelerated multi-direct-forcing immersed boundary method}


\author[SU_staff]{Kosuke Suzuki} 

\author[UE]{Emmanouil Falagkaris} 

\author[UE]{Timm Kr\"{u}ger} 

\author[KU_staff]{Takaji Inamuro} 

\affiliation[SU_staff]{organization={Institute of Engineering, Academic Assembly, Shinshu University},
            city={Nagano},
            postcode={380-8553}, 
            country={Japan}}

\affiliation[UE]{organization={School of Engineering, University of Edinburgh},
            city={Edinburgh},
            postcode={EH9 3FB}, 
            country={Scotland, UK}}
            
\affiliation[KU_staff]{organization={Graduate School of Engineering, Kyoto University},
            city={Kyoto},
            postcode={615-8540}, 
            country={Japan}}
\begin{abstract}
 The multi-direct-forcing immersed boundary method allows for {a small velocity error of} the no-slip condition in moving-particle problems but suffers from numerical instability if simulation parameters are not carefully chosen.
 This study investigates the {boundary-velocity error and numerical} stability of the accelerated multi-direct-forcing immersed boundary method. 
 An analysis of the discretized equations of body motion in moving boundary problems identifies a critical parameter that solely determines the numerical stability for the body motion.
 Additionally, numerical simulations reveal the optimal acceleration parameter that minimizes the {velocity error of} the no-slip condition and is independent of details of the boundary discretisation, the boundary shape, and spatial dimensionality.
 This study provides a guideline for establishing numerically stable simulations of moving boundary problems at optimal {boundary-velocity error}.
\end{abstract}




\end{frontmatter}




\section{Introduction}
One of the key objectives in computational fluid dynamics is to simulate moving boundary problems
with high efficiency and accuracy. 
In recent years, the immersed boundary method (IBM), originally introduced by Peskin in the 1970s to model blood {flows} in the heart~\cite{Peskin1972, Peskin1977}, 
has gained renewed attention as a practical approach for simulating such flows using a fixed Cartesian grid for the fluid phase.

The IBM conceptualizes the boundary as an infinitesimally thin layer, 
filling both the interior and exterior of the boundary with a viscous and usually incompressible fluid.
The no-slip condition on the boundary is enforced by applying a localized volume force near the boundary. 
The core of the IBM lies in how this volume force is determined, and various IBM formulations have been developed based on different strategies for this determination, 
as summarized by Mittal and Iaccarino~\cite{Mittal2005}, Griffith and Patanker~\cite{Griffith2020}, and Verzicco~\cite{Verzicco2023}.

When the boundary represents the surface of a rigid body, 
a widely used variant of IBM is the direct-forcing method~\cite{Mohd1997}. 
In this approach, the boundary is discretized into a series of points, referred to as boundary points. 
The volume force at each boundary point is calculated as the difference between the temporary velocity, which is obtained without considering boundary effects, and the target velocity prescribed by the no-slip condition. 
This computed force is then distributed to the grid points near the boundary. 
There are a number of variations of the direct-forcing IBM, 
and these can be classified into two categories~\cite{Mittal2005}, i.e., sharp-interface schemes and diffuse-interface schemes,
depending on how the boundary points are positioned and the spatial range over which the volume force is distributed to the grid points.
While sharp-interface schemes, such as cut-cell~\cite{Ye1999} and ghost-cell methods~\cite{Mittal2008} can sharply represent the boundary and accurately treat the boundary layer, 
we focus on diffuse-interface schemes due to their simpler algorithms and more straightforward applicability to moving boundary problems.

In a diffuse-interface scheme, the boundary points are treated as Lagrangian points that move along with the rigid body, 
allowing them to be defined independently of the background grid.
The volume force is distributed to the surrounding grid points using a weighting function, 
typically a regularized delta function, as originally proposed by Peskin~\cite{Peskin2002}.
Uhlmann~\cite{Uhlmann2005} was the first to introduce a direct-forcing IBM based on this concept.
Since the algorithm of a diffuse-interface scheme is not tied to the boundary's shape or whether it is stationary or moving,
the diffuse-interface scheme provides a straightforward framework for handling complex moving boundary problems.
However, it generally suffers from {a large velocity error of} the no-slip condition due to the required velocity interpolation and force spreading procedures.

To {reduce the velocity error of} the no-slip condition in the diffuse-interface direct-forcing IBM, 
Wang et al.~\cite{Wang2008} proposed the multi-direct-forcing IBM which iteratively calculates the volume force. 
This method successfully reduced the {boundary-velocity error} by an order of magnitude after ten iterations, but at a higher computational cost.
Wu and Shu~\cite{Wu2009} formulated a matrix equation determining the volume force and proposed the implicit velocity-correction-based IBM which directly solves the matrix equation. 
This method can reduce the {boundary-velocity error} to machine accuracy, but solving the matrix equation greatly increases the computational cost, in particular for moving boundary problems where a large matrix needs to be inverted every time step.
Remarkably, the multi-direct-forcing IBM is equivalent to the Richardson iteration for solving the matrix equation in the implicit velocity-correction-based IBM, 
and its convergence can be accelerated by selecting a suitable acceleration parameter for the Richardson iteration~\cite{Hu2014,Zhang2020} (section~\ref{sec:method} details this consideration).

The present study focuses on the \textit{accelerated} multi-direct-forcing IBM with a single acceleration parameter.
Zhang et al.~\cite{Zhang2020} were the first to propose the accelerated multi-direct-forcing IBM with an optimal acceleration parameter
and investigated the influences of several factors, such as the shape of the boundary and the type of the weighting function, on the acceleration parameter. 
Gsell et al.~\cite{Gsell2019,Gsell2021} proposed a simple correction of the direct-forcing IBM
and successfully reduced the {boundary-velocity error}, just by multiplying the volume force with an error factor which depends on the type of the weighting function.
It turns out that this error factor is equivalent to the acceleration parameter.
The relatively recently introduced accelerated multi-direct-forcing IBM has not been comprehensively investigated.
Especially, the influence of the acceleration parameter on the numerical stability has not been clarified.
In the present study, we investigate not only the {boundary-velocity error} of the accelerated multi-direct-forcing IBM, but also its numerical stability for moving boundary problems. 
Importantly, this study presents simple guidelines for establishing stable simulations of moving boundary problems using the accelerated multi-direct-forcing IBM.

The remainder of this paper is structured as follows: 
Section~\ref{sec:method} outlines the accelerated multi-direct-forcing IBM. 
Section~\ref{sec:stability} discusses its numerical stability for moving boundary problems. 
Section~\ref{sec:result} summarizes the results for stationary and moving boundary problems and discusses the {boundary-velocity error} and stability of the accelerated multi-direct forcing IBM. 
Finally, section~\ref{sec:conclusion} concludes the study.

For a broader discussion, we consider the accelerated multi-direct-forcing IBM without specifying a particular flow solver in section~\ref{sec:method} and section~\ref{sec:stability}. 
However, in section~\ref{sec:result}, we employ the lattice Boltzmann method (LBM)~\cite{Kruger2017, Inamuro2021} as flow solver.
The LBM has been developed into an alternative and promising numerical scheme for simulating viscous fluid flows on the Cartesian grid without solving the Poisson equation for the pressure field. 
Due to its simplicity, computational efficiency, and high scalability in parallel processing, 
the LBM has been widely applied to various simulations, including those involving moving boundaries.
Since both the IBM and LBM operate on a Cartesian grid, the LBM combined with the IBM (referred to as the IB-LBM) is particularly well suited for simulating moving boundary problems. 
Given these advantages, this study supposes that the accelerated multi-direct-forcing IBM is implemented within the IB-LBM framework. 
However, the results and discussion presented here should remain valid even if another flow solver is used.

\section{Accelerated multi-direct-forcing immersed boundary method} \label{sec:method}

\subsection{General remarks on the direct-forcing IBM}

We consider a system where a rigid body moves in an incompressible viscous fluid with density $\rho_{\rm f}$ and dynamic viscosity $\mu$.
The surface of the body is regarded as the boundary for the fluid.
The flow is governed by the incompressible Navier-Stokes equations,
\begin{align}
 \bm{\nabla \cdot u} &= 0, \label{eq:cont} \\
 \rho_{\rm f} \left[\frac{\partial \bm{u}}{\partial t} + (\bm{u \cdot \nabla}) \bm{u}\right] &= - \bm{\nabla} p + \mu \bm{\nabla}^2 \bm{u} + \bm{g}, \label{eq:NS}
\end{align}
where $\bm{u}$ is the flow velocity, $p$ is the pressure, and $\bm{g}$ is the volume force.

\begin{figure}[!tb]
\centering
\includegraphics[width=8cm,clip]{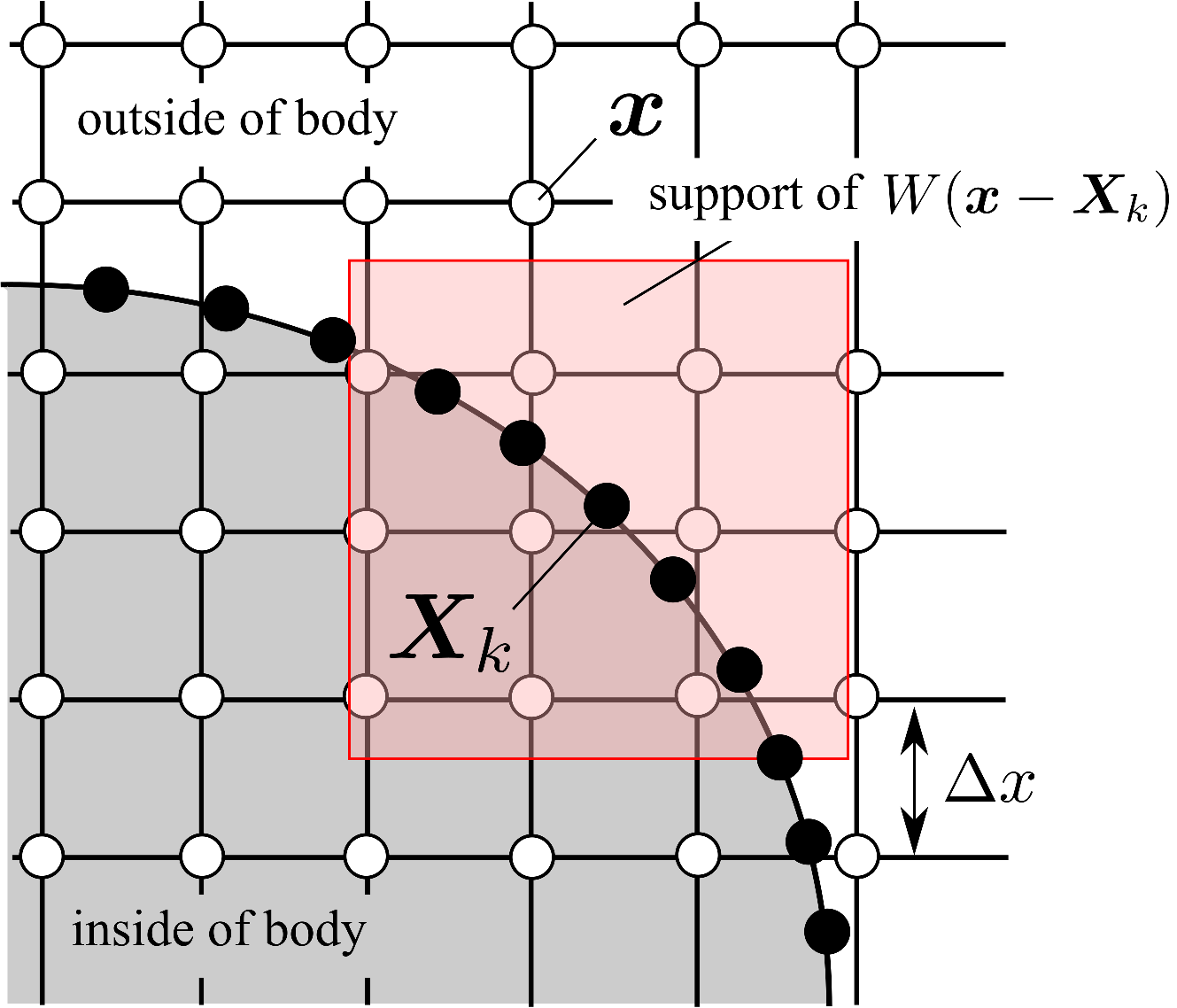} 
\caption{Spatial relationship between the boundary point $\bm{X}_k$ and lattice point $\bm{x}$
in the direct-forcing IBM.
The support of the weighting function $W$ is given by a few lattice spacings $\Delta{x}$. }
\label{fig:ibm}
\end{figure}

In general, the direct-forcing IBM is a method for numerically satisfying the no-slip condition on the boundary.
Suppose that the physical space is divided into a square lattice or a cubic lattice with lattice spacing $\Delta{x}$ and the boundary is divided into $N$ discrete points.
We denote lattice points by $\bm{x}$ and boundary points by $\bm{X}_k$ ($k=1,\ldots, N$) as shown in Fig.~\ref{fig:ibm}.
The direct-forcing IBM is based on the fractional-step approach in computing the incompressible Navier-Stokes equations (\ref{eq:cont}) and (\ref{eq:NS}) as follows:

\begin{enumerate}
 \item Calculate the temporary flow velocity $\bm{u}^*$ by updating the flow velocity from time $t$ to $t+\Delta{t}$ without volume force:
 \begin{equation}
  \bm{u}^*(\bm{x}, t+\Delta{t}) = \bm{u}(\bm{x}, t) + \Delta{t} \left[ - (\bm{u \cdot \nabla} \bm{u}) - \frac{1}{\rho_{\rm f}} \bm{\nabla} p + \frac{\mu}{\rho_{\rm f}} \bm{\nabla}^2 \bm{u} \right] (\bm{x}, t), \label{eq:uast}
 \end{equation}
 where $p$ is determined such that $\bm{u}^*$ satisfies the continuity equation (\ref{eq:cont}).
 In the present study, we suppose that $\bm{u}^*$ is calculated by using the LBM as shown in \ref{sec:algorithm}.
 \item Correct the temporary flow velocity $\bm{u}^*$ by the volume force $\bm{g}$:
 \begin{equation}
  \bm{u}(\bm{x}, t+\Delta{t}) = \bm{u}^*(\bm{x}, t+\Delta{t}) + \frac{\Delta{t}}{\rho_{\rm f}} \bm{g}(\bm{x}, t+\Delta{t}). \label{eq:frac2}
 \end{equation}
\end{enumerate}

The no-slip condition on boundary points at time $t+\Delta{t}$ is represented by
\begin{equation}
\bm{u}(\bm{X}_k, t+\Delta{t}) = \bm{U}_k(t+\Delta{t}), \label{eq:nonslip}
\end{equation}
where $\bm{U}_k$ is the velocity of the boundary point. 
The objective of the direct-forcing IBM is to determine the volume force $\bm{g}(\bm{x}, t+\Delta{t})$ such that it satisfies the no-slip condition (\ref{eq:nonslip}).
{It is noted that in the sharp-interface direct-forcing IBM such as the ghost-cell method~\cite{Mittal2008},
the flow velocity and pressure at lattice points inside the body (namely ghost cells) are directly specified so as to satisfy the no-slip condition.
This procedure is equivalent to applying the volume force $\bm{g}(\bm{x}, t+\Delta{t})$ only within ghost cells. 
Thus, the above formulation provides a general description of direct-forcing IBMs, encompassing both sharp-interface and diffuse-interface schemes.}

To determine the volume force $\bm{g}(\bm{x}, t+\Delta{t})$, 
we require velocity interpolation from $\bm{x}$ to $\bm{X}_k$ and force spreading from $\bm{X}_k$ to $\bm{x}$.
The diffuse-interface direct-forcing IBM assumes that the fluid fills both the outside and inside of the body
and utilizes a weighting function $W$ (whose support is a few lattice spacings, e.g., Fig.~\ref{fig:ibm}) for the interpolation and spreading as follows:
\begin{align}
 \bm{u}^*(\bm{X}_k, t+\Delta{t}) &= \sum_{\bm{x}} \bm{u}^*(\bm{x},t+\Delta{t}) W(\bm{x}-\bm{X}_k) (\Delta{x})^d, \label{eq:interpolate} \\
 \bm{g}(\bm{x}, t+\Delta{t}) &= \sum_{k=1}^N \bm{g}(\bm{X}_k,t+\Delta{t}) W(\bm{x}-\bm{X}_k) \Delta{V}_k, \label{eq:extrapolate}
\end{align}
where $\sum_{\bm{x}}$ denotes summation over all lattice points $\bm{x}$, $d$ represents the spatial dimension ($d=2$ and $3$ in the two- and three-dimensional cases, respectively),
and $\Delta{V}_k$ is a small volume element around the boundary point at the position $\bm{X}_k$ and should satisfy $\sum_{k=1}^N \Delta{V}_k = S \Delta{x}$ (where $S$ is the surface area of the body).
Notably, for the weighting function $W$, Peskin's discrete delta function~\cite{Peskin2002} whose support is $4\Delta{x}$ or $3\Delta{x}$ is often used (see also section~\ref{sec:matrix}).

By substituting Eq.~(\ref{eq:extrapolate}) into Eq.~(\ref{eq:frac2}), performing the interpolation (\ref{eq:interpolate}) for the flow velocities, and then considering the no-slip condition (\ref{eq:nonslip}), 
we have the matrix equation
\begin{equation}
 \mathcal{A} \mathcal{G} = \mathcal{U}, \label{eq:linear}
\end{equation}
where $\mathcal{A}$ is an $N \times N$ matrix whose $(k,l)$ element is given by
\begin{equation}
 \mathcal{A}_{kl} = \sum_{\bm{x}} W(\bm{x}-\bm{X}_k) W(\bm{x}-\bm{X}_l)\Delta{V}_l (\Delta{x})^d, \label{eq:A}
\end{equation}
and $\mathcal{G}$ and $\mathcal{U}$ are $N \times d$ matrices whose $(k, \alpha)$ elements are, respectively, given by
\begin{align}
 \mathcal{G}_{k \alpha} &= g_\alpha (\bm{X}_k, t+\Delta{t}), \\
 \mathcal{U}_{k \alpha} &= \rho_{\rm f} \frac{U_{k \alpha} (t+\Delta{t})- {u}_{\alpha}^*(\bm{X}_k, t+\Delta{t})}{\Delta{t}}.
\end{align}
Therefore, the goal of the diffusive-interface direct-forcing IBM is to calculate $\mathcal{G}$ by solving the matrix equation~(\ref{eq:linear}).
Note that Eq.~(\ref{eq:linear}) is equivalent to the matrix equation for the velocity-correction term in the implicit velocity-correction-based IBM~\cite{Wu2009}.

\subsection{Characteristics of the matrix $\mathcal{A}$} \label{sec:matrix}

To solve the matrix equation~(\ref{eq:linear}), we should know the characteristics of the matrix $\mathcal{A}$ given by Eq.~(\ref{eq:A}).
The matrix $\mathcal{A}$ is determined by the weighting function $W$ and the arrangement of the boundary points $\bm{X}_k$. 
Here, we consider Peskin's discrete delta function~\cite{Peskin2002} whose support is $4\Delta{x}$ or $3\Delta{x}$.
The weighting function $W$ is given by 
\begin{equation}
 W(x,y)=\frac{1}{\Delta{x}}w\left( \frac{x}{\Delta{x}} \right) \cdot \frac{1}{\Delta{x}}w\left( \frac{y}{\Delta{x}} \right) \label{eq:w2}
\end{equation}
in two dimensions and by
\begin{equation}
 W(x,y,z)=\frac{1}{\Delta{x}}w\left( \frac{x}{\Delta{x}} \right) \cdot \frac{1}{\Delta{x}}w\left( \frac{y}{\Delta{x}} \right) \cdot \frac{1}{\Delta{x}}w\left( \frac{z}{\Delta{x}} \right) \label{eq:w3},
\end{equation}
in three dimensions, where
\begin{equation}
 w(r)=\phi_4(r) = \begin{cases}
  {\scriptstyle \frac{1}{8}\left( 3-2|r|+\sqrt{1+4|r|-4r^2}\right) }, & |r|\le 1,  \\
  {\scriptstyle \frac{1}{8}\left( 5-2|r|-\sqrt{-7+12|r|-4r^2}\right) }, & 1\le |r|\le 2,  \\
  {\scriptstyle 0 }, &  \text{otherwise},
 \end{cases}
\end{equation}
or
\begin{equation}
 w(r)= \phi_3(r) = \begin{cases}
  {\scriptstyle \frac{1}{3}\left( 1+\sqrt{1-3r^2}\right) }, & |r|\le 0.5, \\
  {\scriptstyle \frac{1}{6}\left( 5-3|r|-\sqrt{1-3(1-|r|)^2}\right) }, & \ 0.5\le |r|\le 1.5  \\
  {\scriptstyle 0 }, &  \text{otherwise}.
 \end{cases}
\end{equation}
Notably, the weighting function $W$ has a narrow support, making the matrix $\mathcal{A}$ sparse.

As a typical example in two dimensions, we consider the characteristics of $\mathcal{A}$ for a circular cylinder which is represented by evenly distributed boundary points.
We set the diameter of the circular cylinder by $D=50\Delta{x}$
and adjust the relative distance $\Delta{s}=S/(N\Delta{x})$ ($S=\pi D$ in this case) between the neighboring boundary points
by changing the total number $N$ of the boundary points.
The small volume element around each boundary point is given by $\Delta{V}_k = \Delta{s} \Delta{x}$.

\begin{figure}[!tb]
\centering
\includegraphics[width=8cm,clip]{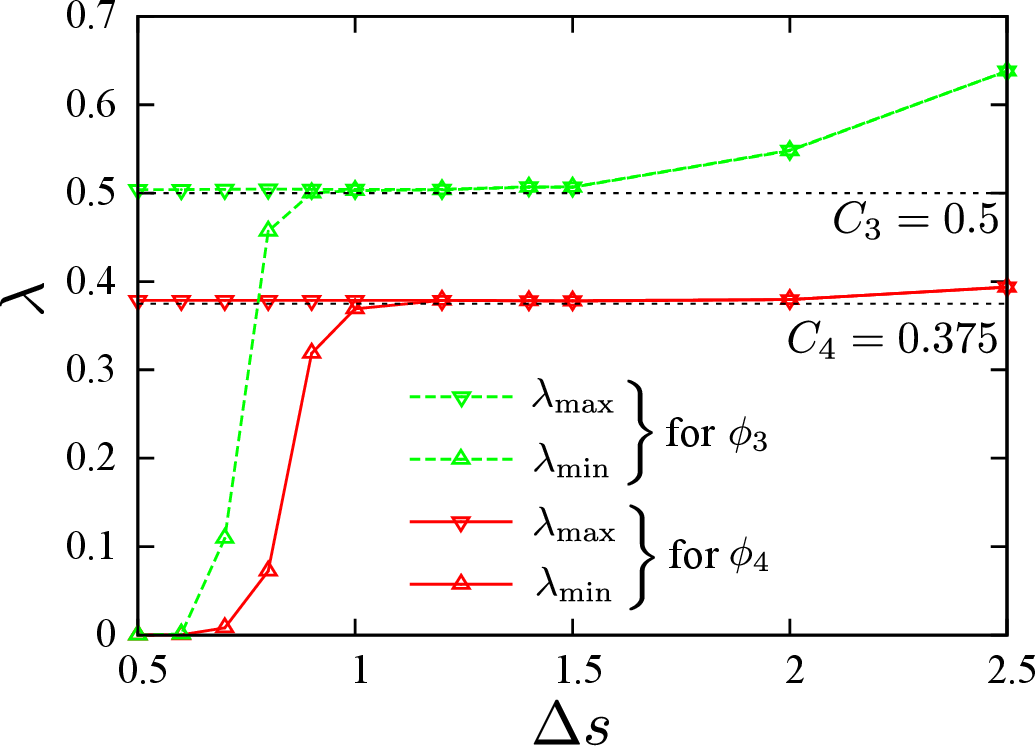} 
\caption{Maximum and minimum eigenvalues of the matrix $\mathcal{A}$ with the weighting function $\phi_3$ or $\phi_4$
for a circular cylinder with various values of $\Delta{s}=S/(N\Delta{x})$.}
\label{fig:matrixA}
\end{figure}

In general, the maximum and minimum eigenvalues provide important information of the matrix $\mathcal{A}$, e.g., the spectral radii of $\mathcal{A}$ and $\mathcal{A}^{-1}$.
We calculate the maximum and minimum eigenvalues of $\mathcal{A}$ for the circular cylinder. 
Fig.~\ref{fig:matrixA} shows the maximum and minimum eigenvalues of $\mathcal{A}$ for various values of $\Delta{s}$.
From this figure, we can see that the maximum eigenvalue $\lambda_{\max}$ is approximately equal to $0.375$ for $\phi_4$ and $0.5$ for $\phi_3$,
when $\Delta{s}$ is smaller than the half width of the support of the weighting function, {i.e., $\Delta{s} \le s/2$}.
For $\Delta{s}$ larger than this width, $\lambda_{\max}$ increases with $\Delta{s}$.
On the other hand, the minimum eigenvalue $\lambda_{\min}$ is approximately equal to $\lambda_{\max}$ for $\Delta{s} \ge 1$, 
but it steeply decreases for $\Delta{s} \le 1$ and tends to $0$.
When $\lambda_{\min}$ is close to $0$, the determinant of $\mathcal{A}$ is also close to $0$, making $\mathcal{A}$ ill-conditioned for small $\Delta{s}$.

Interestingly, the maximum eigenvalues $\lambda_{\max}$ for $\phi_4$ and $\phi_3$ are very close to the constants $C_4 = 0.375$ and $C_3 = 0.5$, respectively,
which are given by
\begin{align}
 C_4 = \sum_j (\phi_4(r-j))^2,~~~C_3 = \sum_j (\phi_3(r-j))^2, \label{eq:constant}
\end{align}
where $\sum_j$ denotes summation over integer values of $j$ and the constants $C_4$ and $C_3$ are independent of $r$.
Eq.~(\ref{eq:constant}) is one of Peskin's conditions that $\phi_4$ and $\phi_3$ must satisfy~\cite{Peskin2002}.

\begin{figure}[!tb]
\centering
\includegraphics[width=8cm,clip]{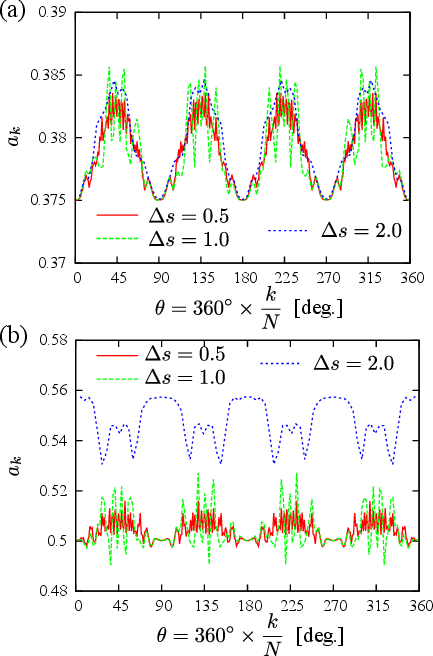} 
\caption{Local value $a_k$ of the norm for the boundary point $\bm{X}_k$
(whose argument is $\theta = 360^\circ \times k/N$ on the circular cylinder)
with the weighting function (a) $\phi_4$ and (b) $\phi_3$ for various values of $\Delta{s}=S/(N\Delta{x})$.}
\label{fig:normA}
\end{figure}

We calculate the maximum norm of $\mathcal{A}$ for the circular cylinder. 
The maximum norm is defined by
\begin{equation}
 \left\|\mathcal{A}\right\|_{\infty} = \max_{1\le k \le N} \sum_{l=1}^N \left|\mathcal{A}_{kl}\right|,
\end{equation}
and it generally satisfies $\left\|\mathcal{A}\right\|_{\infty} \ge \lambda_{\max}$.
To see the relationship between $\left\|\mathcal{A}\right\|_{\infty}$ and $\lambda_{\max}$ in more detail, 
we calculate the \textit{local} value of the norm for each boundary point defined by 
\begin{equation}
 a_k = \sum_{l=1}^N \mathcal{A}_{kl},
\end{equation}
where $\left\|\mathcal{A}\right\|_{\infty} = \max_{1\le k \le N} a_k$, since $\mathcal{A}_{kl} \ge 0$ for all $k$ and $l$.
Fig.~\ref{fig:normA} shows the distribution of $a_k$ with the weighting function $\phi_4$ or $\phi_3$ for different values of $\Delta{s}$.
Evidently, $a_k$ is close to the value of $\lambda_{\max}$ along the surface of the cylinder,
and its relative deviation from $\lambda_{\max}$ is at most $5\%$.
This figure suggests that $\left\|\mathcal{A}\right\|_{\infty}$, as well as $a_k$, are well approximated by $\lambda_{\max}$, independently of the weighting function and the distance between boundary points.

\subsection{Iterative method to solve Eq.~(\ref{eq:linear})} \label{sec:itr}

To directly solve the matrix equation~(\ref{eq:linear}), we have to calculate the inverse of the matrix $\mathcal{A}$ in the same way as the implicit velocity-correction-based IBM~\cite{Wu2009}.
However, the computational cost of calculating $\mathcal{A}^{-1}$ is high due to the large size of $\mathcal{A}$ ($N \times N$), 
especially in moving boundary problems where $\mathcal{A}^{-1}$ must be calculated every time step.
Furthermore, there is a severe restriction in the distance between the boundary points for numerical stability, 
since the steep decrease in $\lambda_{\min}$ for $\Delta{s} \le 1$ makes $\mathcal{A}$ ill-conditioned, as shown in section~\ref{sec:matrix}.
Thus, instead of calculating $\mathcal{A}^{-1}$, several iterative methods have been proposed to solve Eq.~(\ref{eq:linear}), 
and such iterative methods are called multi-direct-forcing IBMs.

The accelerated multi-direct-forcing IBM can be expressed by the Richardson iteration~\cite{Hu2014,Zhang2020} as
\begin{equation}
 \mathcal{G}^{\ell +1} = (\mathcal{I} - \omega \mathcal{A})\mathcal{G}^{\ell} + \omega \mathcal{U},
\end{equation}
where $\mathcal{I}$ denotes the $N \times N$ identity matrix, $\omega$ is the acceleration parameter, 
and the superscript $\ell = 1, 2, 3, \ldots$ is the iteration count.

There are three choices of the acceleration parameter $\omega$ based on previous research~\cite{Wang2008,Zhang2020,Gsell2019,Gsell2021}:
\begin{numcases}{\omega = }
 1, \label{eq:conventional} \\
 C_s^{-1}, \label{eq:Gsell} \\
 \left\|\mathcal{A}\right\|_{\infty}^{-1}, \label{eq:Zhang}
\end{numcases}
where the constant $C_s$ ($s=4$ or $3$) is given by Eq.~(\ref{eq:constant}).
The first choice in Eq.~(\ref{eq:conventional}) corresponds to the conventional multi-direct-forcing IBM~\cite{Wang2008}.
The second choice in Eq.~(\ref{eq:Gsell}) is based on the error factor derived by decomposing the summation of Eq.~(\ref{eq:A}) into the two components which are locally parallel and normal to the boundary~\cite{Gsell2019, Gsell2021}.
The third choice in Eq.~(\ref{eq:Zhang}) is the optimal relaxation parameter of the Richardson iteration~\cite{Zhang2020}.
{Also, we can find a similarity to the feedback IBM proposed by Goldstein et al.~\cite{Goldstein1993} by interpreting $-\omega \rho_{\rm f} / \Delta{t}$ as the feedback coefficient $\beta$.}

As mentioned in section~\ref{sec:matrix}, $\left\|\mathcal{A}\right\|_{\infty} \simeq \lambda_{\max}$, and $\lambda_{\max} \simeq C_s$ for $\Delta{s} \le s/2$.
Thus, it is expected that the choices in Eq.~(\ref{eq:Gsell}) and in Eq.~(\ref{eq:Zhang}) result in similar {boundary-velocity error}. 
In the following, we investigate the {boundary-velocity error} and stability of the IBM for various values of the acceleration parameter $\omega$.

\section{{Stability} in moving boundary problems} \label{sec:stability}

In this section, we discuss the stability for moving boundary problems where a solid body (density $\rho_{\rm c}$, volume $V$) freely moves in a fluid (density $\rho_{\rm f}$).
The full algorithm is shown in \ref{sec:algorithm}.

\subsection{Equation of motion of the body}

Let $\Omega$ represent the closed domain inside the body immersed in the flow field.
Suppose that the volume force $\bm{g}$ is applied to $L_\epsilon$ which represents a narrow neighborhood of the boundary of $\Omega$.
When we calculate the fluid force acting on the body from the volume force of the IBM,
the equation of motion of the body is given by
\begin{equation}
 \rho_{\rm c}V \frac{\text{d}\bm{U}_{\rm c}}{\text{d}t} = -\int_{L_\epsilon} \bm{g}~\text{d}V + \frac{\text{d}}{\text{d}t}\int_{\Omega} \rho_{\rm f} \bm{u}~\text{d}V + (\rho_{\rm c}-\rho_{\rm f})V\bm{G},
\end{equation}
where $\bm{U}_{\rm c}$ is the velocity of the center of mass of the body, 
$\bm{g}$ is the volume force of the IBM, $\bm{u}$ is the flow velocity, and $\bm{G}$ is the gravitational acceleration. 
The internal mass effect (the second term of the right-hand side) is the time derivative of the linear momentum of the fluid inside $\Omega$
and is equal to that of the rigid body with the same density as the fluid~\cite{Suzuki2011}:
\begin{equation}
 \rho_{\rm c}V \frac{\text{d}\bm{U}_{\rm c}}{\text{d}t} = -\int_{L_\epsilon} \bm{g}~\text{d}V + \rho_{\rm f}V \frac{\text{d}\bm{U}_{\rm c}}{\text{d}t} + (\rho_{\rm c}-\rho_{\rm f})V\bm{G}. \label{eq:eom}
\end{equation}

We discretize Eq.~(\ref{eq:eom}) in time and space as follows:
\begin{align}
 \rho_{\rm c}V \frac{\bm{U}_{\rm c}^{n+1}-\bm{U}_{\rm c}^n}{{\Delta{t}}} &= -\sum_{\bm{x}} \bm{g}^n(\bm{x}) (\Delta{x})^d + \rho_{\rm f}V \frac{\bm{U}_{\rm c}^{n}-\bm{U}_{\rm c}^{n-1}}{{\Delta{t}}} + (\rho_{\rm c}-\rho_{\rm f})V\bm{G}, 
\end{align}
where the superscript $n$ is the time step and $\bm{x}$ denotes the lattice points.
In the above equation, the internal mass effect is discretized by using Feng's rigid body approximation~\cite{Feng2009}.
When we use the relation $\sum_{\bm{x}} \bm{g}^n(\bm{x}) (\Delta{x})^d = \sum_{k=1}^{N}\bm{g}^n(\bm{X}_k) \Delta{V}_k$, 
we have
\begin{align}
 \rho_{\rm c}V \frac{\bm{U}_{\rm c}^{n+1}-\bm{U}_{\rm c}^n}{{\Delta{t}}} &= \underbrace{-\sum_{k=1}^{N}\bm{g}^n(\bm{X}_k) \Delta{V}_k}_{\bm{F}^n} + \rho_{\rm f}V \frac{\bm{U}_{\rm c}^{n}-\bm{U}_{\rm c}^{n-1}}{{\Delta{t}}} + (\rho_{\rm c}-\rho_{\rm f})V\bm{G}. \label{eq:govern}
\end{align}
Finally, we have the following difference equation:
\begin{align}
 \bm{U}_{\rm c}^{n+1}-\bm{U}_{\rm c}^n = \frac{\rho_{\rm f}}{\rho_{\rm c}}\left( \bm{U}_{\rm c}^{n}-\bm{U}_{\rm c}^{n-1} \right) + \frac{1}{\rho_{\rm c}V}\bm{F}^n{\Delta{t}} + \left( 1-\frac{\rho_{\rm f}}{\rho_{\rm c}} \right)\bm{G}{\Delta{t}}. \label{eq:difference1}
\end{align}

\subsection{Volume force} \label{sec:volumeforce}

In the multi-direct-forcing IBM, the volume force $\bm{g}^n(\bm{X}_k)$ (and then the fluid force $\bm{F}^n$) is iteratively calculated.
When we consider the case without iteration (i.e., $\ell = 1$), the volume force is given by 
\begin{equation}
 \bm{g}^n(\bm{X}_k) = \omega \rho_{\rm f} \frac{\bm{U}_k^n - \bm{u}^*(\bm{X}_k)}{\Delta{t}}.
\end{equation}
Thus, 
\begin{align}
\bm{F}^n = - \sum_{k=1}^{N} \bm{g}^n(\bm{X}_k) \Delta{V}_k &= - \omega \rho_{\rm f} \sum_{k=1}^{N} \frac{\bm{U}_k^n - \bm{u}^*(\bm{X}_k)}{{\Delta{t}}} \Delta{V}_k \nonumber \\
&= -\omega \rho_{\rm f} \frac{S \Delta{x}}{N}\sum_{k=1}^{N}\frac{\bm{U}_k^n - \bm{u}^*(\bm{X}_k)}{{\Delta{t}}}, \label{eq:gsum}
\end{align}
where we assume $\Delta{V}_k=\frac{S}{N} \Delta{x}$.

When we consider the case with iteration (iteration count $\ell > 1$), 
the volume force $\bm{F}^n_\ell$ is well approximated by a constant multiplication of the volume force $\bm{F}^n_1$ without iteration ($\ell = 1$):
\begin{equation}
 \bm{F}^n = \bm{F}^n_\ell \simeq \eta \bm{F}_1^n = - \eta \omega \rho_{\rm f} \frac{S \Delta{x}}{N}\sum_{k=1}^{N}\frac{\bm{U}_k^n - \bm{u}^*(\bm{X}_k)}{{\Delta{t}}}, \label{eq:gsum2}
\end{equation}
where the constant $\eta$ is derived in \ref{sec:iteration}.
Notably, Eq.~(\ref{eq:gsum2}) includes Eq.~(\ref{eq:gsum}) since $\eta=1$ for $\ell=1$.

\subsection{Discussion on stability} \label{sec:stab}

By substituting Eq.~(\ref{eq:gsum2}) into Eq.~(\ref{eq:difference1}), we have
\begin{align}
 \bm{U}_{\rm c}^{n+1}-\bm{U}_{\rm c}^n = \frac{\rho_{\rm f}}{\rho_{\rm c}}\left( \bm{U}_{\rm c}^{n}-\bm{U}_{\rm c}^{n-1} \right) - \eta \omega \frac{\rho_{\rm f}}{\rho_{\rm c}}\frac{S \Delta{x}}{V} \frac{1}{N} \sum_{k=1}^N \left[ \bm{U}_k^n - \bm{u}^*(\bm{X}_k) \right] + \left( 1-\frac{\rho_{\rm f}}{\rho_{\rm c}} \right)\bm{G}{\Delta{t}}. \label{eq:difference2}
\end{align}
From the coefficients in Eq.~(\ref{eq:difference2}), we can expect that the following factors should affect the numerical stability:
\begin{align}
\gamma &= \frac{\rho_{\rm c}}{\rho_{\rm f}}, \\
A &= \omega \frac{\rho_{\rm f}}{\rho_{\rm c}}\frac{S \Delta{x}}{V} =  \frac{\omega}{\gamma}\frac{S \Delta{x}}{V}. \label{eq:limit}
\end{align}
By using these notations, we have
\begin{align}
\bm{U}_{\rm c}^{n+1}-\bm{U}_{\rm c}^n = \frac{1}{\gamma} \left( \bm{U}_{\rm c}^{n}-\bm{U}_{\rm c}^{n-1} \right) 
- \eta A \frac{1}{N} \sum_{k=1}^N \left[ \bm{U}_k^n - \bm{u}^*(\bm{X}_k) \right] + \left( 1-\frac{1}{\gamma} \right)\bm{G}{\Delta{t}}. \label{eq:difference2_itr}
\end{align}
Since the second term on the right-hand side is determined not only by the translational and angular velocities of the body but also by the flow velocity, 
we cannot evaluate it further from the information of the body velocity $\bm{U}_{\rm c}^n$.
From this equation, however, the numerical stability for moving boundary problems is expected to be determined by $\eta$, $A$, and $\gamma$.

Although the above discussion is based on the equations of translational motion, 
the same analysis can be applied for the equations of rotational motion, 
indicating that the parameters $\eta$, $A$, and $\gamma$ are important in the numerical stability for the overall body motion.
Since the effect of the iteration is confined to the factor $\eta$ multiplying $A$ as shown in Eq.~(\ref{eq:difference2_itr}), 
the product $\eta A$ should be more important than these individual factors in the case with iteration.
We cannot see from Eq.~(\ref{eq:difference2_itr}) whether $\eta A$ or $\gamma$ is more important, 
but in the present study, we found that $\eta A$ dominates the numerical stability of the body motion.
In section~\ref{sec:moving}, we will demonstrate the importance of the parameter ${\eta A}$ through numerical tests of moving boundary problems.

\section{{Numerical simulations} and discussion} \label{sec:result}

In this section, we investigate the {velocity error of} the no-slip condition through simulations of stationary boundary problems
and the stability of the body motion through simulations of moving boundary problems.
In the present simulations, we calculate the temporary flow velocity $\bm{u}^*$ in Eq.~(\ref{eq:uast}) by using the lattice Boltzmann method as shown in \ref{sec:algorithm}.
{It should be noted that although the density in the LBM exhibits small variations from the constant fluid density $\rho_{\rm f}$ due to weak compressibility,
the flow velocity and pressure calculated by the LBM converge to the incompressible Navier-Stokes equations (\ref{eq:cont}) and (\ref{eq:NS}) with the second-order accuracy~\cite{Kruger2017,Inamuro2021},
and the density variation is a negligible error which does not affect the second-order accuracy.}

\subsection{Stationary boundary problems}

\subsubsection{Circular cylinder fixed in a Poiseuille flow} \label{sec:circular_fix}

First, we examine the {velocity error of} the no-slip condition
on the surface of a circular cylinder fixed in a Poiseuille flow.
The domain size is $L\times H = 200\Delta x \times 200 \Delta{x}$, 
and the top and bottom boundaries of the domain are no-slip walls.
The left and right boundaries are periodic with an imposed pressure difference $\Delta{p}$~\cite{Inamuro2021}, 
inducing the Poiseuille flow (in the absence of the cylinder) with the mean flow velocity
\begin{equation}
 U = \frac{H^2}{12\rho_{\rm f} \nu}\frac{\Delta{p}}{L}, \label{eq:char_U}
\end{equation}
where $\nu$ is the kinematic viscosity.
A circular cylinder with diameter $D=50\Delta{x}$ is fixed at the center of the domain,
and it is represented by evenly distributed boundary points.
We define the Reynolds number as
\begin{equation}
 Re = \frac{UH}{\nu}. \label{eq:Re}
\end{equation}
In the present simulations, we set $Re = 100$.
Other parameters are set as $\rho_{\rm f} = 1$, 
$\nu = 0.06~(\Delta{x})^2/\Delta{t}$, $\Delta{p} = 1.08 \times 10^{-4}~(\Delta{x}/\Delta{t})^2$, and $U=0.03~\Delta{x}/\Delta{t}$ in simulation units.
We run the simulations until the non-dimensional time $t^* = tU/H = 75$, by which the flow field has reached a steady state and the {boundary-velocity error} remains unchanged.

\begin{figure}[!tb]
\centering
\includegraphics[width=8cm]{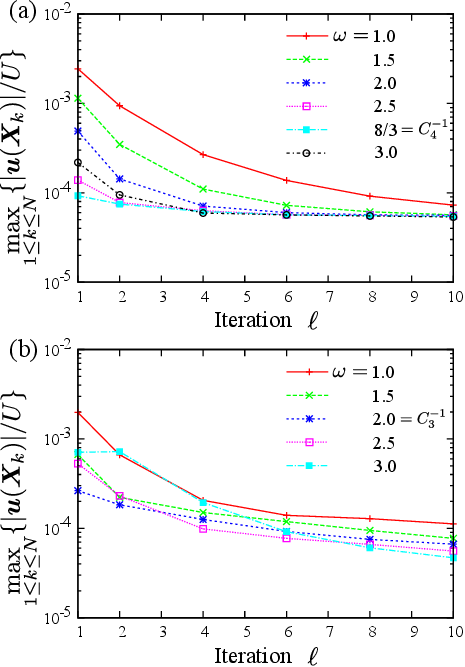} 
\caption{Maximum {boundary-velocity error} on the circular cylinder as a function of the iteration count $\ell$
with the weighting function (a) $\phi_4$ and (b) $\phi_3$ for various values of the acceleration parameter $\omega$.}
\label{fig:circ_itr}
\end{figure}

Fig.~\ref{fig:circ_itr} shows the maximum {boundary-velocity error} on the circular cylinder with $\Delta{s} \simeq 1.0$ ($N=157$) in steady state.
For the weighting function $\phi_4$, the error decreases monotonically with the number of iterations and significantly depends on the acceleration parameter $\omega$.
Notably, the error for $\omega = C_4^{-1}$ is about ten times as small as that for $\omega = 1.0$ with $\ell = 1$, 
indicating that a good choice of $\omega$ can significantly {reduce the boundary-velocity error} even without iteration. 
This trend can also be observed for the weighting function $\phi_3$, although {the boundary-velocity error} for $\phi_3$ and $\omega = C_{3}^{-1}$ is inferior to that for $\phi_4$ and $\omega = C_{4}^{-1}$.

\begin{figure}[!tb]
\centering
\includegraphics[width=8cm]{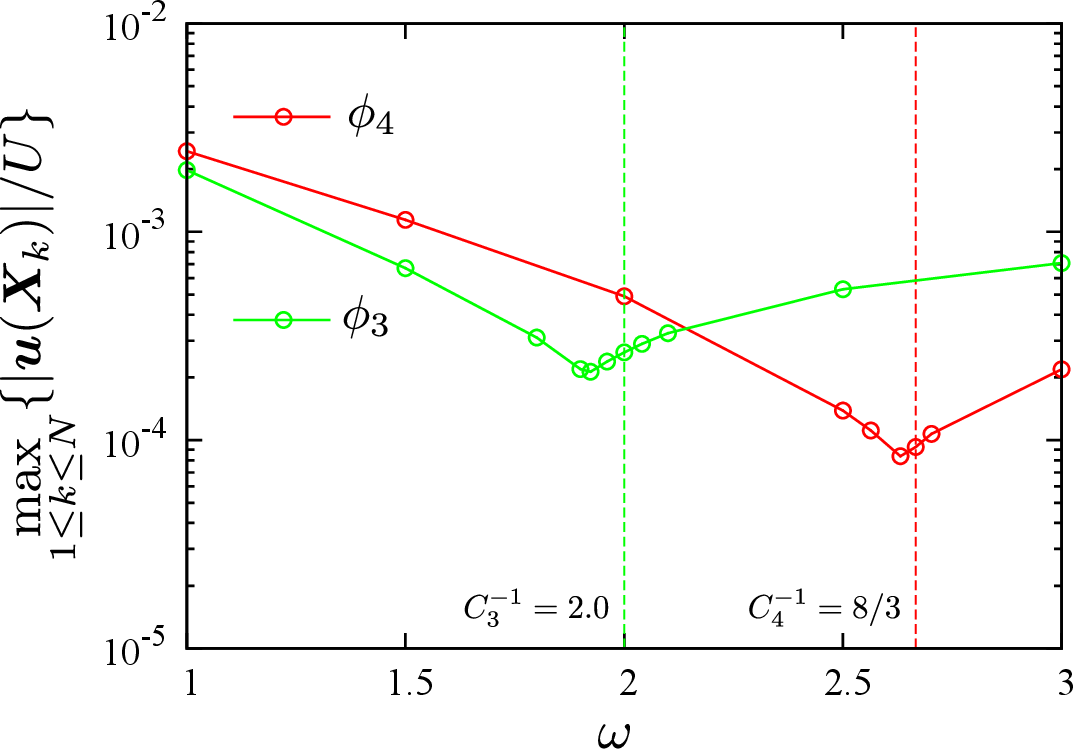} 
\caption{Maximum {boundary-velocity error} on the circular cylinder as a function of the acceleration parameter $\omega$
with the weighting function $\phi_4$ and $\phi_3$ without iteration ($\ell = 1$).}
\label{fig:err_vs_omega}
\end{figure}

In order to find the optimal value of $\omega$, we plot the error without iteration ($\ell = 1$) as a function of $\omega$ as shown in Fig.~\ref{fig:err_vs_omega}.
From this figure, we can see that the optimal values of $\omega$ for $\phi_4$ and $\phi_3$ are approximately equal to $C_4^{-1}=8/3$ and $C_3^{-1} = 2$, respectively. 
The maximum eigenvalues of the matrix $\mathcal{A}$ for $\phi_4$ and $\phi_3$ are $\lambda_{\rm max} = 0.3787$ and $0.5042$, respectively, 
and the maximum norms of $\mathcal{A}$ for $\phi_4$ and $\phi_3$ are $\left\|\mathcal{A}\right\|_{\infty} = 0.3857$ and $0.5270$, respectively.
Thus, $C_s^{-1} \simeq \lambda_{\rm max}^{-1} \simeq \left\|\mathcal{A}\right\|_{\infty}^{-1}$ for both $s=4$ and $s=3$, 
and $\omega = C_{s}^{-1}$ is almost the optimal choice to minimize the {boundary-velocity error}. 
These findings are consistent with our expectation in section~\ref{sec:itr} that the choices (\ref{eq:Gsell}) and (\ref{eq:Zhang}) should result in the same {boundary-velocity error}.

\begin{figure}[!tb]
\centering
\includegraphics[width=8cm,clip]{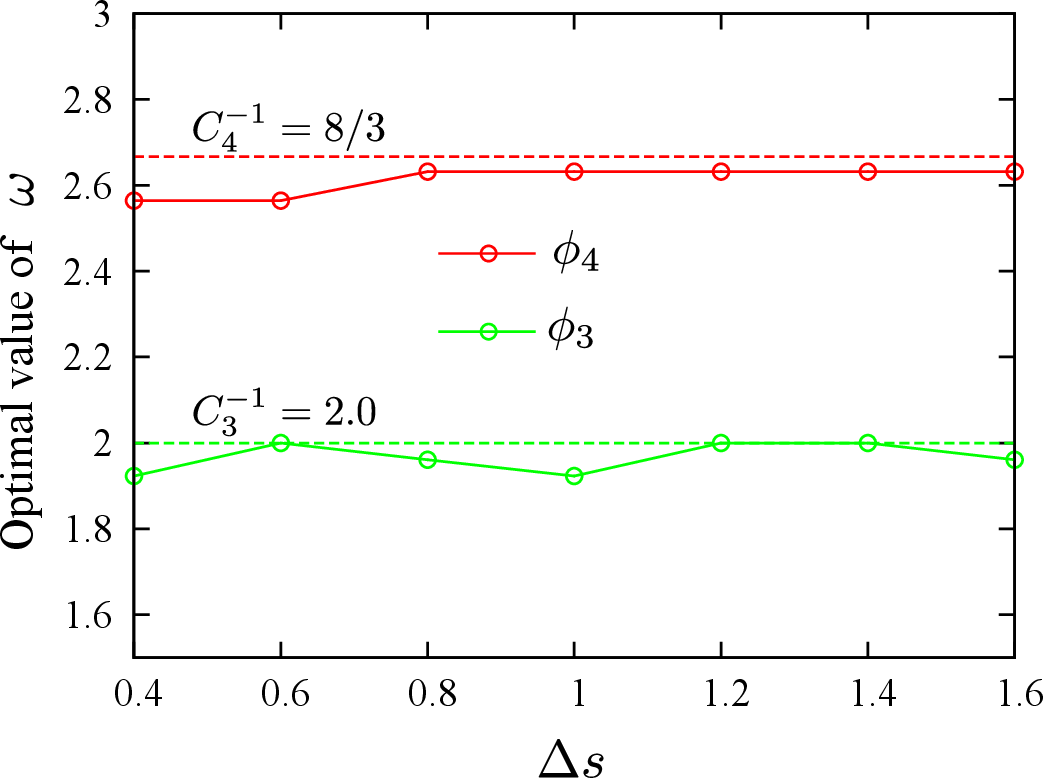} 
\caption{Optimal values of the acceleration parameter $\omega$
for the weighting functions $\phi_4$ and $\phi_3$ without iteration ($\ell = 1$)
for the circular cylinder with various values of $\Delta{s}=S/(N\Delta{x})$.}
\label{fig:opt_ds}
\end{figure}

Fig.~\ref{fig:opt_ds} shows the optimal value of $\omega$ as a function of the distance $\Delta{s}$ between the neighboring boundary points.
This figure indicates that the optimal value of $\omega$ is approximately equal to $C_{s}^{-1}$, independently of $\Delta{s}$
in the range of $0.4 \le \Delta{s} \le 1.6$.
Notably, the simulations for $\omega \simeq C_{s}^{-1}$ are stable in this range of $\Delta{s}$, 
which includes the range where $\mathcal{A}$ becomes ill-conditioned, 
and the maximum {boundary-velocity errors} have comparable values for various values of $\Delta{s}$. 
Therefore, the choice of $\omega = C_{s}^{-1}$ is almost optimal in terms of the {velocity error of} the no-slip condition independently of $\Delta{s}$.

\begin{figure}[!tb]
\centering
\includegraphics[width=10cm,clip]{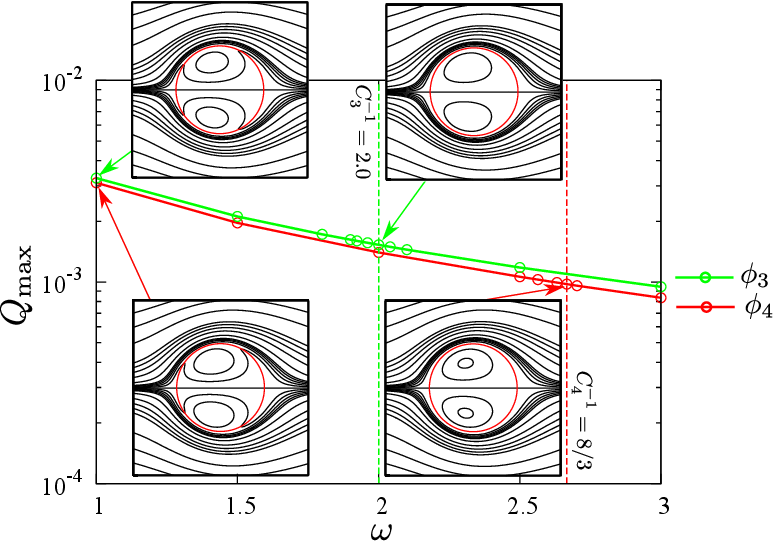} 
\caption{Penetration magnitude $Q_{\rm max}$ across the surface of the circular cylinder with $\Delta{s} = 1.0$ as a function of the acceleration parameter $\omega$
for the weighting functions $\phi_4$ and $\phi_3$ without iteration ($\ell = 1$); 
the insets display the streamlines around the circular cylinder.}
\label{fig:penet_omega}
\end{figure}

From another viewpoint, we examine the penetration across the surface of the circular cylinder, 
which has negative impacts on the flow field around the boundary as well as the force acting on the boundary.
In order to evaluate the penetration, we define the penetration magnitude $Q_{\rm max}$ as follows:
\begin{equation}
Q_{\rm max} = \frac{\max_{1\le k \le N} \{ |\psi(\bm{X}_k)-\psi(\bm{X}_{k+1})| \}}{U S/N},
\end{equation}
where $\psi$ is the stream function defined by $u=\partial \psi /\partial y$ and $v= - \partial \psi /\partial x$.
We calculate $\psi$ at lattice points by integrating $u$ and $v$ using the trapezoidal rule
and then interpolate it at boundary points $\bm{X}_k$ (we set $\bm{X}_{N+1} = \bm{X}_{1}$ here).
In general, $|\psi(\bm{X})-\psi(\bm{Y})|$ is equal to the volume flow rate passing between two points $\bm{X}$ and $\bm{Y}$. 
Thus, the penetration magnitude $Q_{\rm max}$ indicates the maximum volume flow rate passing between neighboring boundary points.
Fig.~\ref{fig:penet_omega} shows the values of $Q_{\rm max}$ as a function of the acceleration parameter $\omega$ for $\Delta{s} = 1.0$ without iteration ($\ell = 1$).
From this figure, we can see that $Q_{\rm max}$ monotonically decreases with increasing $\omega$ both for $\phi_4$ and $\phi_3$.
Notably, unlike the boundary-velocity error shown in Fig.~\ref{fig:err_vs_omega}, $Q_{\rm max}$ does not have its minimum around the optimal value of $\omega$. 
In other words, as $\omega$ increases beyond the optimal value, 
the penetration magnitude becomes smaller while the boundary-velocity error becomes larger.
In the streamlines shown in the insets of Fig.~\ref{fig:penet_omega}, penetration occurs across the surface of the circular cylinder for $\omega = 1.0$, 
whereas penetration does not occur for the optimal value of $\omega$ ($\omega = C^{-1}_4$ for $\phi_4$ and $\omega = C^{-1}_3$ for $\phi_3$). 
Therefore, the optimal choice of $\omega$ not only minimizes the boundary-velocity error but also prevents streamlines from penetrating across the boundary even without iteration.

In summary, the acceleration parameter $\omega$ improves the convergence of the iteration, as shown in Fig.~\ref{fig:circ_itr}.
Furthermore, there exists an optimal value, $\omega \simeq C_s^{-1}$, which minimizes the boundary–velocity error (Fig.~\ref{fig:err_vs_omega}) and prevents streamlines from penetrating across the boundary (Fig.~\ref{fig:penet_omega}).
Remarkably, the present study shows that even without iteration, the optimal choice of $\omega$ yields a boundary-velocity error comparable to that obtained after several iterations in the conventional multi-direct-forcing IBM.
Therefore, setting $\omega = C_s^{-1}$ and $\ell = 1$ is sufficient in most cases, although additional iterations might be used when smaller boundary-velocity errors are desired.

\subsubsection{Elliptical cylinder fixed in a Poiseuille flow} \label{sec:elliptical_fix}

Next, we examine the {velocity error of} the no-slip boundary condition
on the surface of an elliptical cylinder, whose boundary points are distributed unevenly, 
fixed in a Poiseuille flow.
The computational domain and parameters are the same as those in section~\ref{sec:circular_fix}, 
except that an elliptical cylinder (long and short radii are $a=16\Delta{x}$ and $b=8\Delta{x}$, respectively) is located at the center of the domain instead of a circular cylinder. 
The major axis of the elliptical cylinder is inclined by $\alpha = 45^\circ$ with respect to the flow axis.
The locations of the boundary points of the cylinder are given by
\begin{align}
 \bm{X}_k = \left[ \begin{array}{cc}
  \cos \alpha & - \sin \alpha \\
  \sin \alpha & \cos \alpha
 \end{array}\right]\left( \begin{array}{c}
  a \cos \left(2\pi \frac{k}{N} \right) \\
  b \sin \left(2\pi \frac{k}{N} \right)
 \end{array}\right),~~~k=1,2,\ldots , N.
\end{align}
In this arrangement, the boundary points are distributed unevenly, 
and $0.5\Delta{x}^2 \le \Delta{V}_k \le 1.0\Delta{x}^2$ for $N=100$.
The maximum and minimum eigenvalues of the matrix $\mathcal{A}$ are 
$\lambda_{\max} = 0.3798$ and $\lambda_{\min} = 1.280 \times 10^{-4}$, respectively, 
and $\left\|\mathcal{A}\right\|_\infty = 0.3856$.
Thus, $\lambda_{\max} \simeq ||\mathcal{A}||_\infty \simeq C_4 = 0.375$.

\begin{figure}[!tb]
\centering
\includegraphics[width=8cm]{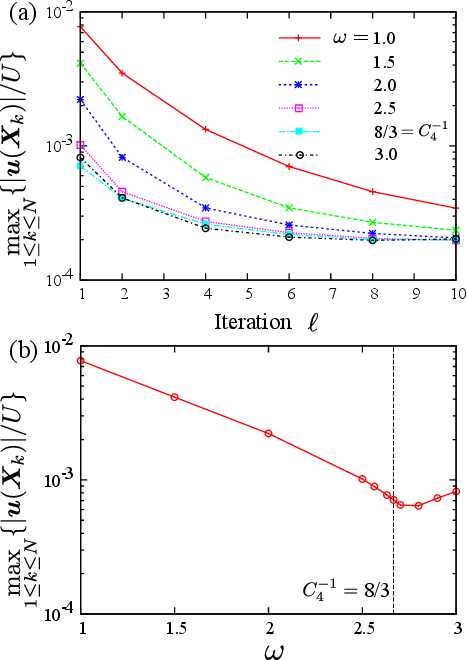} 
\caption{Maximum {boundary-velocity error} on an elliptical cylinder with the weighting function $\phi_4$:
(a) dependence of the iteration count $\ell$ for various values of the acceleration parameter $\omega$; 
(b) dependence of $\omega$ without iteration ($\ell = 1$).}
\label{fig:ellip_itr}
\end{figure}

Fig.~\ref{fig:ellip_itr} shows the maximum {boundary-velocity error} on the elliptical cylinder in steady state for the weighting function $\phi_4$.
From Fig.~\ref{fig:ellip_itr}(a), we can see the same trend as for the circular cylinder in section~\ref{sec:circular_fix}, 
i.e., the error decreases monotonically with the iteration count, and the error for $\omega = C_4^{-1}$ is about ten times as small as that for $\omega = 1.0$ with $\ell = 1$. 
Also, from Fig.~\ref{fig:ellip_itr}(b), we can see that the optimal value of $\omega$ is approximately equal to $C_4^{-1}$ in the same way as for the circular cylinder.
Therefore, even when the boundary points are unevenly spaced, the choice of $\omega = C_{s}^{-1}$ can optimize the {velocity error of} the no-slip condition. 
The weighting function $\phi_3$ gives similar results (not shown).

\subsubsection{Sphere fixed in a planar Poiseuille flow} \label{sec:sphere_Poiseuille_fix}

Finally, we consider the three-dimensional flow around a fixed sphere.
The domain size is $L\times H \times H = 200\Delta x \times 200 \Delta{x}\times 200 \Delta{x}$.
The top and bottom boundaries of the domain (which are perpendicular to the $y$-axis) are the no-slip walls, 
the left and right boundaries (which are perpendicular to the $x$-axis) are periodic with the pressure difference $\Delta{p}$, 
and the other boundaries (which are perpendicular to the $z$-axis) are periodic without a pressure difference.
Thus, this computational condition induces a planar Poiseuille flow similar to that in section~\ref{sec:circular_fix} and section~\ref{sec:elliptical_fix}.
The characteristic flow speed $U$ and the Reynolds number $Re$ are defined by Eq.~(\ref{eq:char_U}) and Eq.~(\ref{eq:Re}).

\begin{figure}[!tb]
\centering
\includegraphics[width=8cm]{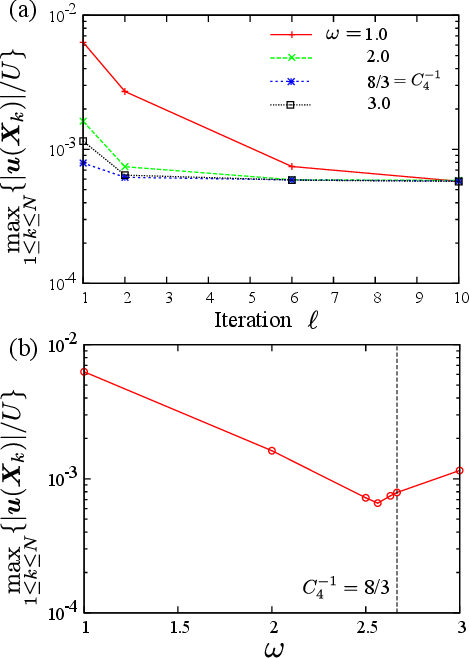} 
\caption{Maximum {boundary-velocity error} on a sphere with the weighting function $\phi_4$:
(a) dependence of the iteration count $\ell$ for various values of the acceleration parameter $\omega$; 
(b) dependence of $\omega$ without iteration ($\ell = 1$).}
\label{fig:sph_itr}
\end{figure}

A sphere with diameter $D=28\Delta{x}$ is fixed at the center of the domain.
The boundary points representing the sphere are almost evenly distributed by giving the points a virtual Coulomb potential
and optimizing the positions of the boundary points so that the total potential is minimized. 
The distance between any two neighboring boundary points is between $0.7939\Delta{x}$ and $1.085\Delta{x}$ with a mean distance of $0.9403\Delta{x}$.
Thus, assuming that $\Delta{V}_k$ is constant, we set $\Delta{V}_k = \frac{S}{N} \Delta{x} = 0.9996\Delta{x}^3$ for $N=2464$.
The maximum and minimum eigenvalues of the matrix $\mathcal{A}$ for the weighting function $\phi_4$ are 
$\lambda_{\max} = 0.3812$ and $\lambda_{\min} = 3.935 \times 10^{-2}$, respectively, 
and $\left\|\mathcal{A}\right\|_\infty = 0.4020$.
Thus, $\lambda_{\max} \simeq \left\|\mathcal{A}\right\|_\infty \simeq C_4 = 0.375$.

Fig.~\ref{fig:sph_itr} shows the maximum {boundary-velocity error} on the sphere in steady state.
From Fig.~\ref{fig:sph_itr}(a), we can see the same trend as for the two-dimensional results in section~\ref{sec:circular_fix} and section~\ref{sec:elliptical_fix}, 
i.e., the error decreases monotonically with the iteration count, and the error for $\omega = C_4^{-1}$ is about ten times as small as that for $\omega = 1.0$ with $\ell = 1$. 
Also, from Fig.~\ref{fig:ellip_itr}(b), we can see that the optimal value of $\omega$ is approximately equal to $C_4^{-1}$, in the same way as for the two-dimensional results.
Therefore, even for three-dimensional problems, the choice of $\omega = C_{s}^{-1}$ can optimize the {velocity error of} the no-slip condition. 
As before, the weighting function $\phi_3$ gives similar results (not shown).

\subsection{Moving boundary problems} \label{sec:moving}

\subsubsection{Circular cylinder moving in a Poiseuille flow} \label{sec:circular}

In order to test the numerical stability discussed in section~\ref{sec:stability}, 
we simulate the motion of a circular cylinder in a Poiseuille flow.
The same computational domain as in section~\ref{sec:circular_fix} is used, i.e., 
the domain size is $L\times H = 200\Delta x \times 200 \Delta{x}$, 
the top and bottom boundaries of the domain are no-slip walls, 
and the left and right boundaries are periodic with an applied pressure difference $\Delta{p}$.
We consider the motion of a circular cylinder with diameter $D$ and density $\rho_{\rm c}$ in this pressure-driven flow.
The fluid and the cylinder are initially at rest and start to be driven by the pressure gradient.
The center of the cylinder is initially located at $(x_{\rm c}, y_{\rm c}) = (0.5L, 0.4H)$, slightly off the channel centerline.
We do not consider gravity, i.e., $\bm{G} = \bm{0}$.
The characteristic flow speed $U$ and the Reynolds number $Re$ are defined in Eq.~(\ref{eq:char_U}) and Eq.~(\ref{eq:Re}), respectively, 
and are set as $U=0.01~\Delta{x}/\Delta{t}$ and $Re = 40$ with $\nu = 0.05~(\Delta{x})^2/\Delta{t}$.
The distance between neighboring boundary points is approximately kept at $S/N = \Delta{x}$ for various values of $D$.
We use $\phi_4$ as the weighting function.
It is noted that in two-dimensional simulations (sections~\ref{sec:circular} and \ref{sec:sed_circular}), all equations of motion, including both translational and rotational motions, are calculated using the first-order Euler method (see also \ref{sec:algorithm}). 

\begin{figure}[!tb]
\centering
\includegraphics[width=8cm]{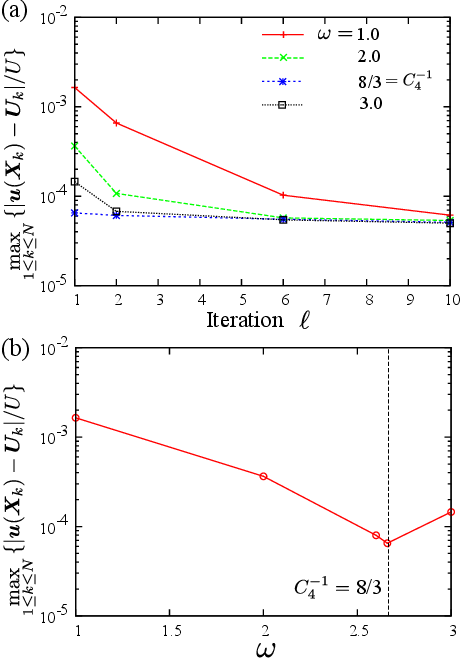} 
\caption{Maximum {boundary-velocity error} on a moving circular cylinder:
(a) dependence of the iteration count $\ell$ for various values of the acceleration parameter $\omega$; 
(b) dependence of $\omega$ without iteration ($\ell = 1$).}
\label{fig:moving_circ_itr}
\end{figure}

First, we check the {velocity error of} the no-slip condition.
We run stable simulations with $D=20\Delta{x}$ and $\gamma = 1.0$ until the non-dimensional time $t^* = tU/H = 50$, by which the linear and angular velocities of the circular cylinder remain almost unchanged.
Since the {boundary-velocity error} oscillates with time even in the steady state, we calculate the time-averaged error over $40 \le t^* \le 50$.
Fig.~\ref{fig:moving_circ_itr} shows the maximum {boundary-velocity error} on the moving circular cylinder.
In the same way as for stationary boundary problems, 
the error decreases monotonically with the iteration count, and the optimal value of $\omega$ is approximately equal to $C_4^{-1}$. 
Thus, the general behavior of the {boundary-velocity error} is independent of stationary or moving boundaries. 

In general, the motion of a moving body with a small density ratio tends to become unstable for weakly coupled algorithms.
The present simulations also become unstable for small values of $\gamma$.
Furthermore, large values of $\omega$ and small values of $D$ also make the simulations unstable.
Thus, there is a range of these parameters for achieving stable simulations of moving boundary problems.
As discussed in section~\ref{sec:stability}, the parameter $A$ given in Eq.~(\ref{eq:limit}) should play an important role in the numerical stability.
Since $S/V = \pi D/(\pi D^2/4) = 4/D$ for the circular cylinder, 
\begin{equation}
 A = \frac{\omega}{\gamma}\frac{4 \Delta{x}}{D}.
\end{equation}
In the following, we modify the value of $A$ by changing $\omega$, $\gamma$, and $\Delta{x}/D$ (which corresponds to the non-dimensional spatial resolution) and examine whether the motion of the circular cylinder results in a blow-up or not.
For a ``blow-up", the force $\bm{F}^n$ starts to oscillate after several ten time steps, 
its amplitude grows in time,
and eventually the velocity of the circular cylinder becomes too large to properly calculate the flow velocity around it,
or the cylinder jumps out of the computational domain, typically within several hundred time steps.

\begin{figure}[!t]
\centering
\includegraphics[width=12cm,clip]{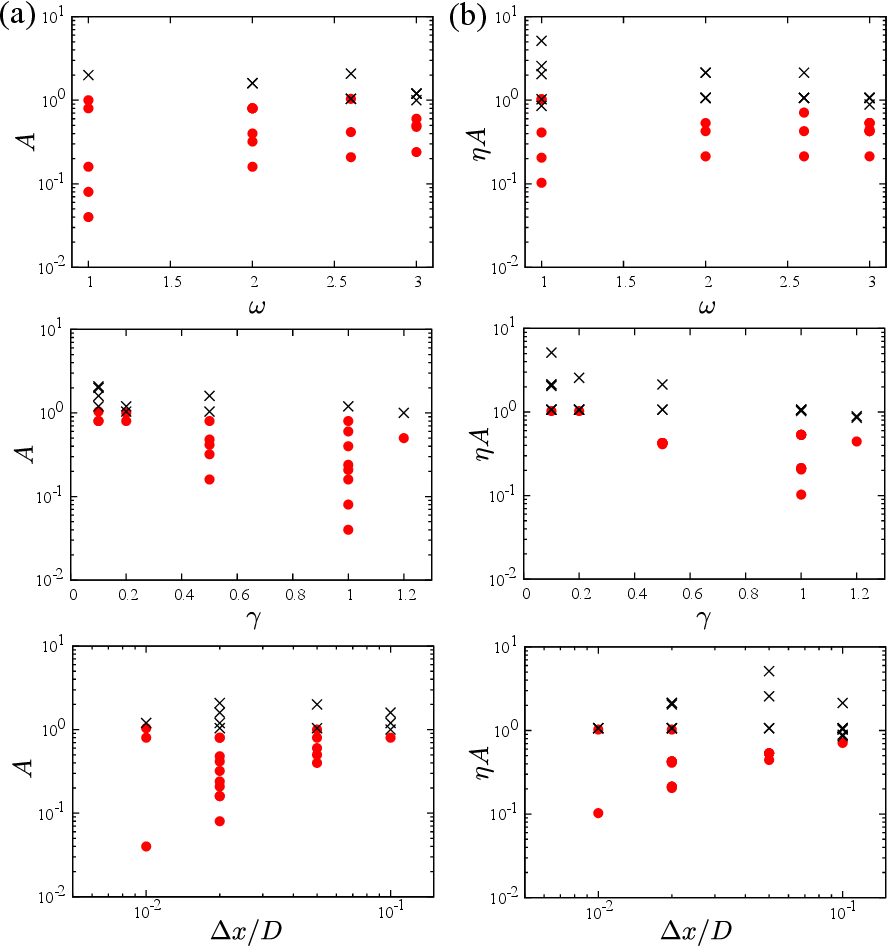} 
\caption{(a) Maps of $\omega$--$A$, $\gamma$--$A$, and $\Delta{x}/D$--$A$
({$\bullet$}: stable, $\times$: unstable)
without iteration ($\ell = 1$) for a circular cylinder moving in a Poiseuille flow; 
(b) maps of $\omega$--$\eta A$, $\gamma$--$\eta A$, and $\Delta{x}/D$--$\eta A$ with iteration ($\ell = 6$).}
\label{fig:mapA_circular}
\end{figure}

Fig.~\ref{fig:mapA_circular}(a) shows the stability maps in the plane of $\omega$--$A$, $\gamma$--$A$, and $\Delta{x}/D$--$A$ without iteration ($\ell = 1$).
From this figure, we can see that there is an upper limit $A \lessapprox 1.0$ for stable simulations,
and the limit value is almost independent of $\omega$, $\gamma$, and $\Delta{x}/D$.
In addition, in our preliminary calculations for 
$(Re, \nu, U, S/N) = (40, 0.25(\Delta{x})^2/\Delta{t}$, $0.05\Delta{x}/\Delta{t}, 1.0\Delta{x})$, 
$(40, 0.05(\Delta{x})^2/\Delta{t}, 0.01\Delta{x}/\Delta{t}, 1.4\Delta{x})$, $(40, 0.05(\Delta{x})^2/\Delta{t}, 0.01\Delta{x}/\Delta{t}, 0.7\Delta{x})$, $(1.0, 2.0(\Delta{x})^2/\Delta{t}, 0.01\Delta{x}/\Delta{t}, 1.0\Delta{x})$, and $(1.0, 0.2(\Delta{x})^2/\Delta{t}, 0.001\Delta{x}/\Delta{t}, 1.0\Delta{x})$, 
the results coincide with the maps in Fig.~\ref{fig:mapA_circular}(a), 
indicating that the limit value is independent of the values of $Re$, $\nu$, $U$, and $S/N$.
The weighting function $\phi_3$ gives the same results.
Thus, at least for the current case, the numerical stability can be determined solely by the value of $A$, 
and there is an upper limit $A \lessapprox 1.0$ for stable simulations.


Goldstein et al.~\cite{Goldstein1993} also suggested a stability limit in their feedback IBM, where the upper limit of the time step $\Delta{t}$ is determined by the feedback coefficients.
This stability limit might seem similar to the upper limit of $A$ in the present study, 
since the acceleration parameter $\omega$ corresponds to the feedback coefficient $\beta$, as mentioned in section~\ref{sec:itr}.
However, they are entirely different from each other. 
The stability limit in the feedback IBM~\cite{Goldstein1993} is for the numerical stability of the flow field around the boundary, 
whereas the upper limit of $A$ in the present study is for the numerical stability of the body motion in moving boundary problems involving the fluid--body interaction.
Although the numerical stability for moving boundary problems has been discussed in previous studies (e.g., \cite{Borazjani2008,Lacis2016}), 
the solid--fluid density ratio $\gamma$ has typically been used as the indicator for determining whether a simulation becomes stable or unstable.
However, the present study demonstrates that the parameter $A$ serves as a more appropriate indicator than the density ratio $\gamma$. 
Therefore, the critical value of $A$ provides a quantitative guideline for the \textit{a priori} choice of simulation parameters leading to numerically stable simulations of moving boundary problems.

It is expected that, in the case with iteration, the numerical stability should be determined by $\eta A$, rather than by $A$, as discussed in section~\ref{sec:stab}.
Fig.~\ref{fig:mapA_circular}(b) shows the results with iteration ($\ell =6$), 
using $\eta A$ instead of $A$ for the vertical axis.
From this figure, we can see that there is an upper limit $\eta A \lessapprox 1.0$ for stable results,
and the limit value is almost independent of $\omega$, $\gamma$, and $\Delta{x}/D$ and close to that in the case without iteration.
{We can obtain the same conclusion for any value of $\ell$ (not shown). 
In the following subsections, we primarily discuss the case without iteration, i.e., $\ell = 1$ and thus $\eta = 1$ (see \ref{sec:iteration}), 
for which the numerical stability is expected to be determined solely by $\eta A = A$.}


\subsubsection{Sedimentation of a circular cylinder} \label{sec:sed_circular}

As the second problem, we consider the sedimentation of a circular cylinder. 
In this case, we examine the numerical stability for various values of the gravitational acceleration $\bm{G}$.
We use the same computational domain and conditions as in the previous problem, except that we set $\Delta{p}=0$ and $\bm{G} = (G,0)$. 
It is noted that we consider gravity acting to the right rather than downward to simplify the description using the previously defined computational domain and conditions.
Hence, we are viewing the system in a coordinate system rotated by $90^{\circ}$, without changing the physical scenario.

\begin{figure}[!t]
\centering
\includegraphics[width=12cm,clip]{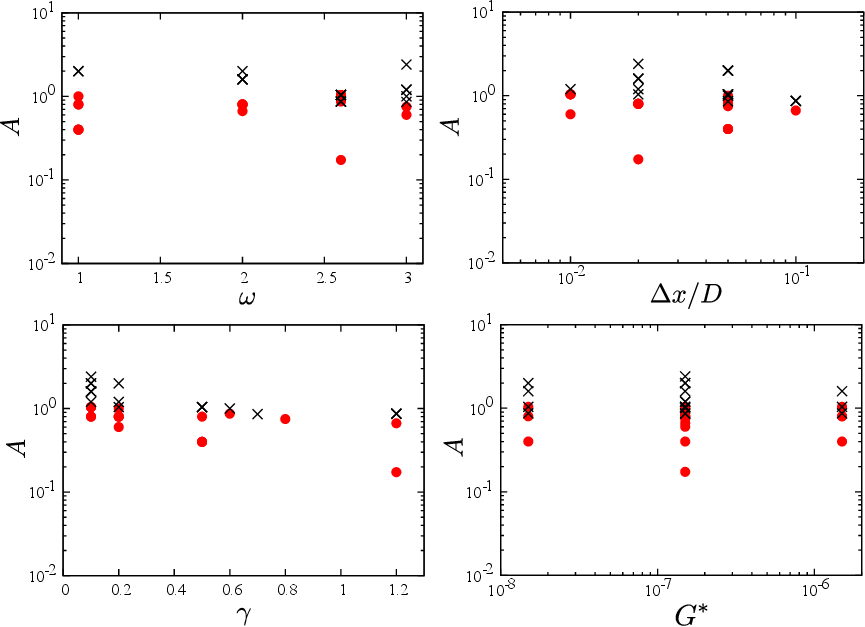} 
\caption{Maps of $\omega$--$A$, $\gamma$--$A$, $\Delta{x}/D$--$A$, and $G^*$--$A$
({$\bullet$}: stable, $\times$: unstable)
without iteration ($\ell = 1$) for the sedimentation of a circular cylinder.}
\label{fig:mapA_circular_sedimentation}
\end{figure}

Fig.~\ref{fig:mapA_circular_sedimentation} shows the stability maps in the plane of $\omega$--$A$, $\gamma$--$A$, $\Delta{x}/D$--$A$, and $G^*$--$A$ (where $G^*=G(\Delta{t})^2/\Delta{x}$) without iteration ($\ell = 1$).
It is noted that the results for $G=0$ or $\gamma=1.0$ are omitted because the circular cylinder does not move.
Although there is a narrow region where stable and unstable results coexist 
(e.g., two different sets of parameters $(\omega, \gamma, \Delta{x}/D, G^*) = (2.6, 0.1, 0.01, 1.5 \cdot 10^{-6})$ and $(2.6, 0.5, 0.05, 1.5 \cdot 10^{-6})$ give the same value of $A = 1.04$, 
but the former is stable while the latter is unstable), 
the results are mostly separated by the parameter $A$.
Thus, the numerical stability does not depend on the value of $G$ alone and is determined solely by the parameter $A$.
The upper limit $A \lessapprox 1.0$ for stable results is comparable to that for the previous problem in section~\ref{sec:circular}.

\subsubsection{Sphere moving in a planar Poiseuille flow} \label{sec:sphere_Poiseuille}

Finally, we simulate the motion of a sphere in a planar Poiseuille flow.
The same computational domain as in section~\ref{sec:sphere_Poiseuille_fix} is used,
and a planar Poiseuille flow is induced by the pressure difference $\Delta{p}$.
A sphere with diameter $D$ and density $\rho_{\rm c}$ moves in this pressure-driven flow.
The fluid and the sphere are initially at rest and start to be driven by the pressure gradient.
The center of the sphere is initially located at $(x_{\rm c}, y_{\rm c}, z_{\rm c}) = (0.5L, 0.4H, 0.5H)$, slightly off the channel centerline.
We do not consider gravity, i.e., $\bm{G} = \bm{0}$.
The characteristic flow speed $U$ and the Reynolds number $Re$ are defined in Eq.~(\ref{eq:char_U}) and Eq.~(\ref{eq:Re}), respectively, 
and are set as $U=0.01~\Delta{x}/\Delta{t}$ and $Re = 40$ with $\nu = 0.05~(\Delta{x})^2/\Delta{t}$.
The boundary points representing the sphere are almost evenly distributed in the same way as in section~\ref{sec:sphere_Poiseuille_fix}, 
and the number $N$ of the boundary points is determined so that $S/N = 1.0 (\Delta{x})^2$.
The distance between the two closest boundary points is kept approximately at $\Delta{x}$.
We use $\phi_4$ as the weighting function.

\begin{figure}[!t]
\centering
\includegraphics[width=6cm,clip]{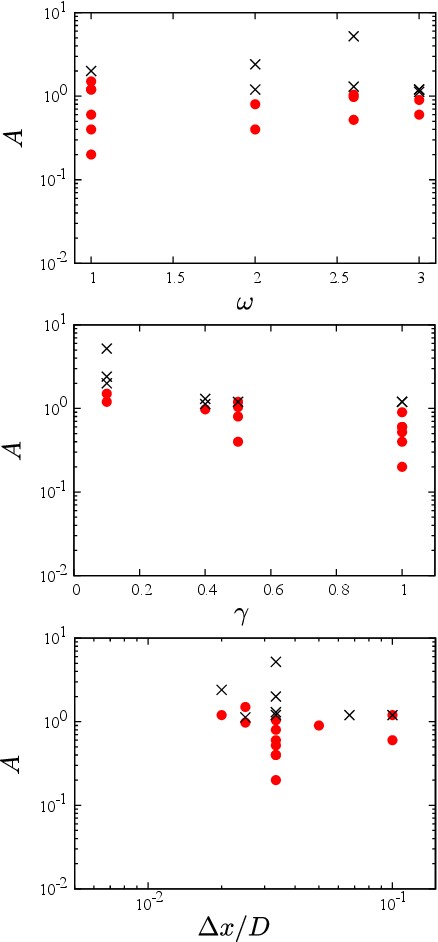} 
\caption{Maps of $\omega$--$A$, $\gamma$--$A$, and $\Delta{x}/D$--$A$ ({$\bullet$}: stable, $\times$: unstable)
without iteration ($\ell = 1$) for a sphere moving in a planar Poiseuille flow.}
\label{fig:mapA_sphere}
\end{figure}

Since $S/V = \pi D^2/(\pi D^3/6) = 6/D$ for the sphere, 
\begin{equation}
 A = \frac{\omega}{\gamma}\frac{6 \Delta{x}}{D}. \label{eq:A_sphere}
\end{equation}
Thus, we modify the value of $A$ by changing $\omega$, $\gamma$, and $\Delta{x}/D$ and examine whether the motion of the sphere results in a blow-up or not.
It is noted that the equation of motion of the sphere is computed by using the first-order Euler method, except the kinematic equations for the rotational motion, 
which is computed by the second-order modified Euler method to satisfy the normalization condition for the quaternions (see Ref.~\cite{Suzuki2011}).

Fig.~\ref{fig:mapA_sphere} shows the stability maps in the plane of $\omega$--$A$, $\gamma$--$A$, and $\Delta{x}/D$--$A$ 
in the case without iteration ($\ell = 1$).
From this figure, we can find a boundary between stable and unstable simulations at $A \simeq 1.0$ for various values of $\omega$, $\gamma$, and $\Delta{x}/D$.
The upper limit of $A$ for stable results is slightly higher than that in the two-dimensional case in section~\ref{sec:circular}, 
presumably because the kinematic equations for the rotational motion are computed by the second-order modified Euler method. 
Thus, even in three-dimensional problems, the numerical stability for moving boundary flows can be determined by the value of $A$, and its limit value is $A \lessapprox 1.0$.

\subsection{Complex case demonstrations}
Finally, we apply the accelerated multi-direct-forcing IBM to more complex problems such as butterfly flight~\cite{Suzuki2024a} and ice slurry flow~\cite{Suzuki2024b},
which have been well validated using the conventional multi-direct-forcing IBM,
in order to demonstrate that the accelerated method produces results comparable to the conventional method while requiring significantly less computational cost.

\subsubsection{Butterfly flight} \label{sec:butterfly}
We consider the same problem as Ref.~\cite{Suzuki2024a}, i.e., free-flight simulations of a butterfly model with wing-pitch flexibility and thorax-pitch control. 
The system of the butterfly model is a four-body problem composed of two rigid flat plates (left and right wings) and two rigid rods (the thorax and abdomen). 
We set the maximum thorax-pitch angle to $\alpha_{\rm m} = 60^\circ$ and the non-dimensional wing-pitch spring stiffness to $N_{\rm K} = 2.00$ as one of the computational parameter sets investigated in Ref.~\cite{Suzuki2024a}.
The reference length is given by $L_{\rm ref} = 120\Delta{x}$, 
the flapping period is given by $T=12000\Delta{t}$, 
and the mean flapping speed is given by $U_{\rm ref} = 0.0314 \Delta{x}/\Delta{t}$. 
{The gravitational acceleration is given by $G =9.99 \cdot 10^{-7} \Delta{x}/(\Delta t)^2$.}
The boundary points are almost evenly distributed on the model, and the number of the boundary points is $N=14668$ for one wing, which gives $\Delta{V}_k = 0.978 \Delta{x}^3$. 
We use $\phi_4$ as the weighting function.
The equations of motion of the butterfly model are described by five independent variables
(position coordinates $x$ and $y$, thorax-pitch angle $\theta_{\rm th}$, 
wing-pitch angle $\theta_{\rm wp}$, and abdominal angle $\theta_{\rm ab}$)
and calculated by using the second-order Adams--Bashforth method. 
For more details of the butterfly model and computational conditions, please see Ref.~\cite{Suzuki2024a}. 

\begin{figure}[!t]
\centering
\includegraphics[width=12cm,clip]{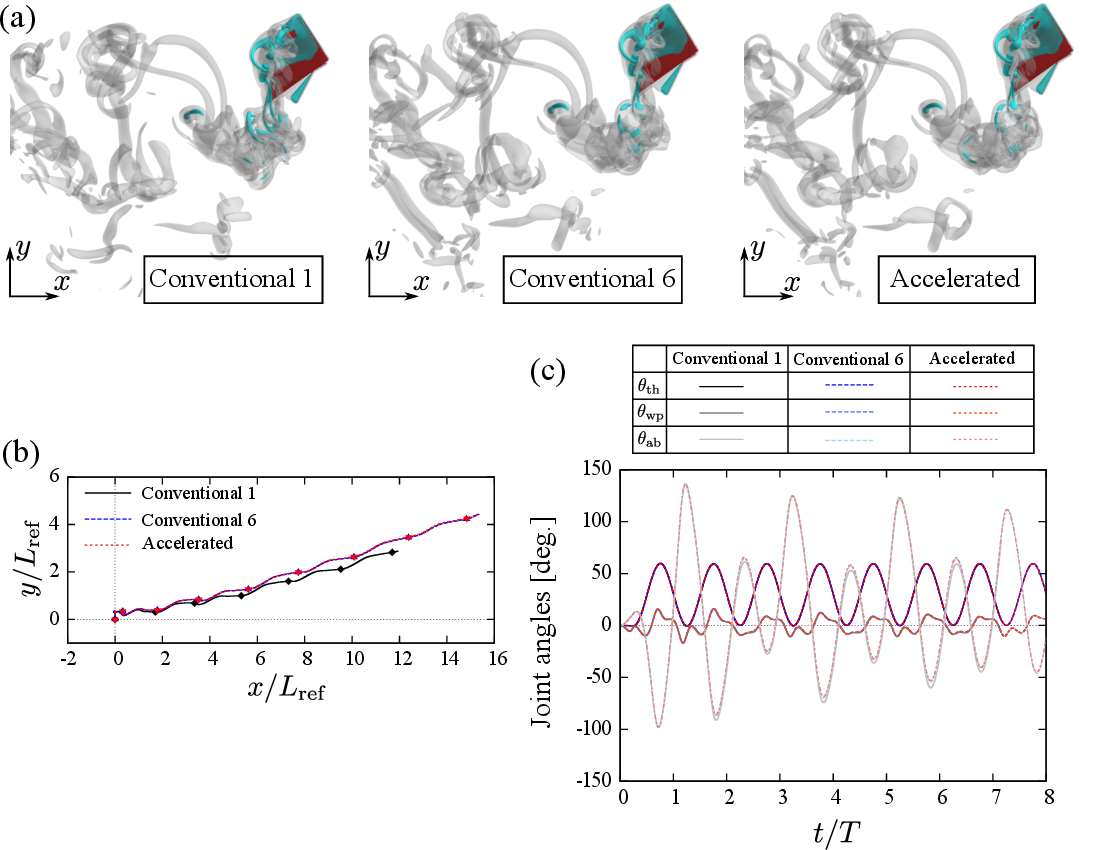} 
\caption{Comparison between the conventional direct-forcing IBM ($\omega = 1.0$, $\ell = 1$),
the conventional multi-direct-forcing IBM ($\omega = 1.0$, $\ell = 6$),
and the accelerated multi-direct-forcing IBM ($\omega=C_{4}^{-1}$, $\ell = 1$) in the simulations of the butterfly flight:
(a) vortex structures viewed from the right (red surface denotes the model surface; grey and blue surfaces denote the isosurface of $|\bm{\nabla} \times \bm{u}| = 5 U_{\rm ref}/L_{\rm ref}$ and $20 U_{\rm ref}/L_{\rm ref}$, respectively) at $t/T = 3.0$; 
(b) trajectories of the center of the thorax (symbols indicate the position when the wings are at top dead point);
(c) time variations of the thorax-pitch angle $\theta_{\rm th}$, wing-pitch angle $\theta_{\rm wp}$, and abdominal angle $\theta_{\rm ab}$. }
\label{fig:butterfly_result}
\end{figure}

We set the acceleration parameter $\omega$ and the iteration count $\ell$ as $(\omega, \ell) = (1.0, 1)$, $(1.0, 6)$, and $(C_{4}^{-1}, 1)$. 
The first set corresponds to the conventional direct-forcing IBM,
the second set corresponds to the conventional multi-direct-forcing IBM with six iterations, 
and the third set corresponds to the accelerated multi-direct-forcing IBM with the optimal acceleration parameter without iteration. 
Hereafter, we refer to the first one as ``Conventional 1," 
to the second one as ``Conventional 6,"
and to the third one as ``Accelerated."
From the results shown in the previous sections, it is expected that the {boundary-velocity error} is comparable between Conventional 6 and Accelerated. 
Fig.~\ref{fig:butterfly_result} shows the comparison between Conventional 1 and 6 and Accelerated in terms of the vortex structures, trajectories, and joint angles of the butterfly model. 
From this figure, we can see that the results for Conventional 6 and Accelerated are almost identical, while Conventional 1 shows a visible difference in the trajectory. 
Thus, Conventional 6 and Accelerated give almost the same results even in a complex problem with non-spherical multiple rigid bodies, while Conventional 1 includes a visible error due to the {large boundary-velocity error}. 
Since the results for Conventional 6 have been extensively validated as mentioned in Ref.~\cite{Suzuki2024a}, Accelerated also gives reasonable results for this problem. 

Table~\ref{tab:comp_butterfly} shows the {velocity error of} the no-slip condition and the computational time. 
In this table, the values of the {boundary-velocity error} are shown as its time-averaged values in $1 \le t/T\le 2$, where we cut off the initial transient data.  
The computational time is measured in $0 \le t/T\le 2$.
These simulations were performed on 24 cores of a Xeon Silver 4214 (2.2 GHz) using Message Passing Interface (MPI).
From this table, we can see that the {boundary-velocity errors} are comparable between Conventional 6 and Accelerated, 
and the {boundary-velocity error} for Conventional 1 is much larger than the errors for Conventional 6 and Accelerated, 
as expected from the results shown in previous sections (e.g., see Fig.~\ref{fig:moving_circ_itr}(a)). 
The computational time for Accelerated is reduced from that for Conventional 6 due to the reduction of the proportion of the IBM. 
This is naturally attributed to the fact that the iteration of the IBM is reduce from $\ell = 6$ to $1$. 
Therefore, Accelerated produces results comparable to Conventional 6 while requiring less computational cost.

\begin{table}[t!]
\begin{center}
\caption{Mean and maximum {boundary-velocity errors}, computational time per time step, and proportion of the LBM, IBM, body motion, and other parts to the total computing time
in the simulations of the butterfly flight for the conventional and accelerated multi-direct-forcing IBMs.}
\label{tab:comp_butterfly}
\scalebox{0.8}[0.8]{
  \begin{tabular}{llccc} \hline
                  & & Conventional 1 & Conventional 6& Accelerated  \\ 
                  & & ($\omega = 1.0$, $\ell = 1$) & ($\omega = 1.0$, $\ell = 6$) & ($\omega = C_{4}^{-1}$, $\ell = 1$)  \\ \hline
  	{Boundary-velocity error} & Mean & $2.169 \times 10^{-2}$ & $2.334 \times 10^{-3}$ & $2.818 \times 10^{-3}$ \\
	              & Maximum & $3.615 \times 10^{-1}$ & $1.156 \times 10^{-1}$ & $1.913 \times 10^{-1}$ \\ \hline
	Computational time (s) & per time step & 5.646 & 7.400 & 5.629  \\
	Proportion (\%) of	& LBM & 87.13 & 66.48 & 87.15 \\ 
					& IBM & 12.75 & 33.43 & 12.73 \\
					& Body motion & \phantom{0}0.01 & \phantom{0}0.01 & \phantom{0}0.01 \\
					& Other & \phantom{0}0.11 & \phantom{0}0.08 & \phantom{0}0.11 \\ \hline
  \end{tabular}}
 \end{center}
\end{table}

In order to consider the numerical stability, we evaluate the parameter $A$ for this problem. 
This system includes multiple rigid bodies, and the part with the smallest mass should determine the numerical stability.
Thus, we should consider the parameter $A$ for the wings instead of that for the whole model. 
Although the wings do not have volume, i.e., $V = 0$, we can regard $\rho_{\rm c} V$ as the wing mass $M_{\rm w}$. 
In addition, the surface area $S$ of the wings is equal to $2L_{\rm ref}^2$.
Thus, the parameter $A$ for the wings is 
\begin{equation}
 A = \omega \rho_{\rm f} \frac{S \Delta{x}}{M_{\rm w}} = \omega \left( \frac{M}{M_{\rm w}} \right) \left( \frac{\rho_{\rm f} L_{\rm ref}^3 }{M} \right) \frac{2 L_{\rm ref}^2 \Delta{x}}{L_{\rm ref}^3} = \frac{\omega}{\WR \, N_{\rm M}}\frac{2\Delta{x}}{L_{\rm ref}}, \label{eq:A_butterfly}
\end{equation}
where $M$ is the mass of the whole model, 
$\WR = M_{\rm w}/M = 0.1$ is the wing-mass ratio, 
and $N_{\rm M} = M/(\rho_{\rm f} L_{\rm ref}^3) = 5.03$ is the non-dimensional mass.
In the simulation using Accelerated, the value of $A$ is $8.836 \times 10^{-2} \ll 1$, which is consistent with the fact that the simulations are numerically stable. 
To find the boundary between stable and unstable simulations for this problem, we change the parameter $A$ by changing the non-dimensional mass $N_{\rm M}$. 
As a result, the motion of the model results in numerical instability for $A \ge 0.635$ ($N_{\rm M} \le 0.7$), 
whereas the motion of the model is numerically stable for $A \le 0.556$ ($N_{\rm M} \ge 0.8$). 
Thus, there is an upper limit of $A$ for stable results in the same way as in the previous sections, although its limit value is slightly less than that for a moving sphere. 
This finding is presumably because the model is composed of multiple rigid bodies with complex shape far from a sphere. 
{A similar phenomenon has been observed in vortex-induced-vibration simulations of a circular cylinder~\cite{Lacis2016},
where adding a splitter plate to the cylinder increases the critical density ratio for weak-coupling algorithms.}

In summary, even in a complex problem with non-spherical multiple rigid bodies, 
the accelerated multi-direct-forcing IBM can give almost the same results as the conventional multi-direct-forcing IBM while reducing the computational cost. 
Furthermore, the numerical stability of the body motion is determined by the value of $A$, although the limit value is slightly different from that for a moving sphere
presumably due to the complex shape of the model.

\subsubsection{Ice slurry flow}
We consider the same problem as Ref.~\cite{Suzuki2024b}, i.e., ice slurry flow within a circular tube.
We calculate the motions of many spherical particles in a fluid and the heat transfer between the particles and the fluid. 
The particles are kept at low temperature $T_{\rm p} = 0$, and the tube wall is kept at high temperature $T_{\rm w} = 1$.
Thus, we calculate not only the flow field but also the temperature field by using the IB-LBM.
The descriptions of the LBM and IBM for the temperature field are omitted here and found in Ref.~\cite{Suzuki2021}.
In this simulation, we set the Reynolds number $Re = 1000$, Grashof number $Gr = 0$ {(i.e., gravity is absent)}, density ratio $\gamma = 1.0$, ice-packing factor $\IPF = 15.0$\%, and diameter ratio $\DR = 0.1$ as one of the computational parameter sets investigated in Ref.~\cite{Suzuki2024b}.
In this case, the particle diameter is $D = 15 \Delta{x}$, 
the tube diameter is $H=150\Delta{x}$, 
the number of the boundary points for one particle is $N=714$ (which gives $\Delta{V}_k = 0.989 \Delta{x}^3$), the number of particles is $N_{\rm p} = 225$, and the cross-sectional-averaged flow speed is $U_{\rm ref} = 0.04 \Delta{x}/\Delta{t}$.
We use $\phi_4$ as the weighting function.
The equations of motions of particles are solved in the same way as in the case of a single moving sphere shown in section~\ref{sec:sphere_Poiseuille},
{although repulsive forces during particle–particle and particle–wall collisions prevent a particle from penetrating into another particle or the tube wall.}
For more details of the system and computational conditions, please see Ref.~\cite{Suzuki2024b}. 

\begin{figure}[!t]
\centering
\includegraphics[width=12cm,clip]{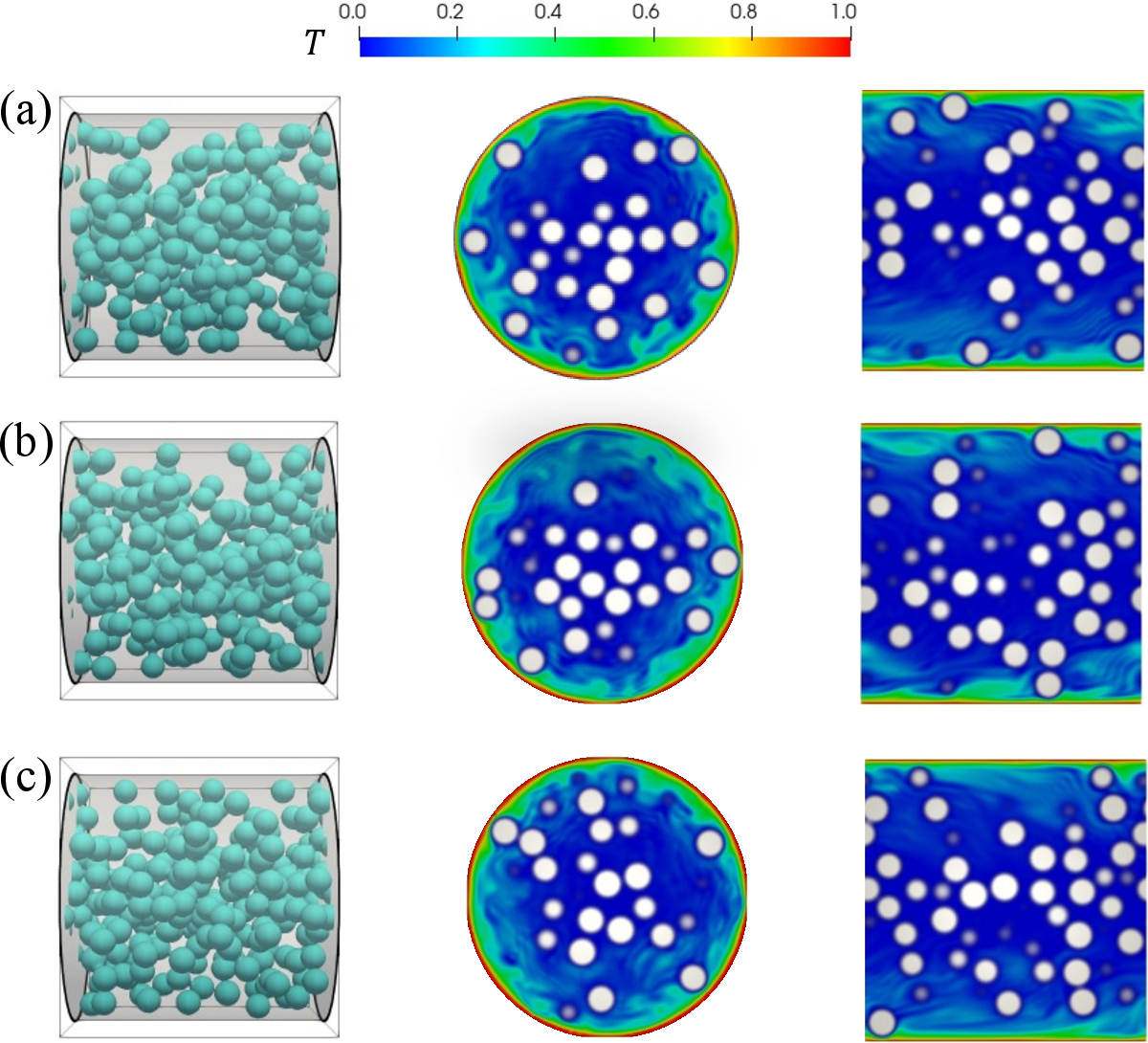} 
\caption{Snapshots captured at $tU_{\rm ref}/H = 140$ depicting particles (left) and color maps of the temperature $T$ viewed from the outlet (center) and from the side (right)
for (a) the conventional direct-forcing IBM ($\omega = 1.0$, $\ell = 1$),
(b) the conventional multi-direct-forcing IBM ($\omega = 1.0$, $\ell = 6$), 
and (c) the accelerated multi-direct-forcing IBM ($\omega=C_{4}^{-1}$, $\ell =1$). }
\label{fig:iceslurry_result1}
\end{figure}

We simulate this problem by using 
Conventional 1 ($\omega = 1.0$, $\ell = 1$), 
Conventional 6 ($\omega = 1.0$, $\ell = 6$), 
and Accelerated ($\omega = C_{4}^{-1}$, $\ell = 1$) in the same way as in section~\ref{sec:butterfly}.
Fig.~\ref{fig:iceslurry_result1} depicts the distributions of particles and the temperature field for Conventional 1 and 6 and Accelerated at the non-dimensional time $t^* = tU_{\rm ref}/H = 140$. 
From this figure, we can see that the particles are homogeneously distributed throughout the circular tube, and the thermal boundary layer on the tube wall is disturbed by the particles. 
While the detailed particle configuration is different, the thermal boundary layer appears comparable in size between Conventional 1 and 6 and Accelerated.

\begin{figure}[!t]
\centering
\includegraphics[width=8cm,clip]{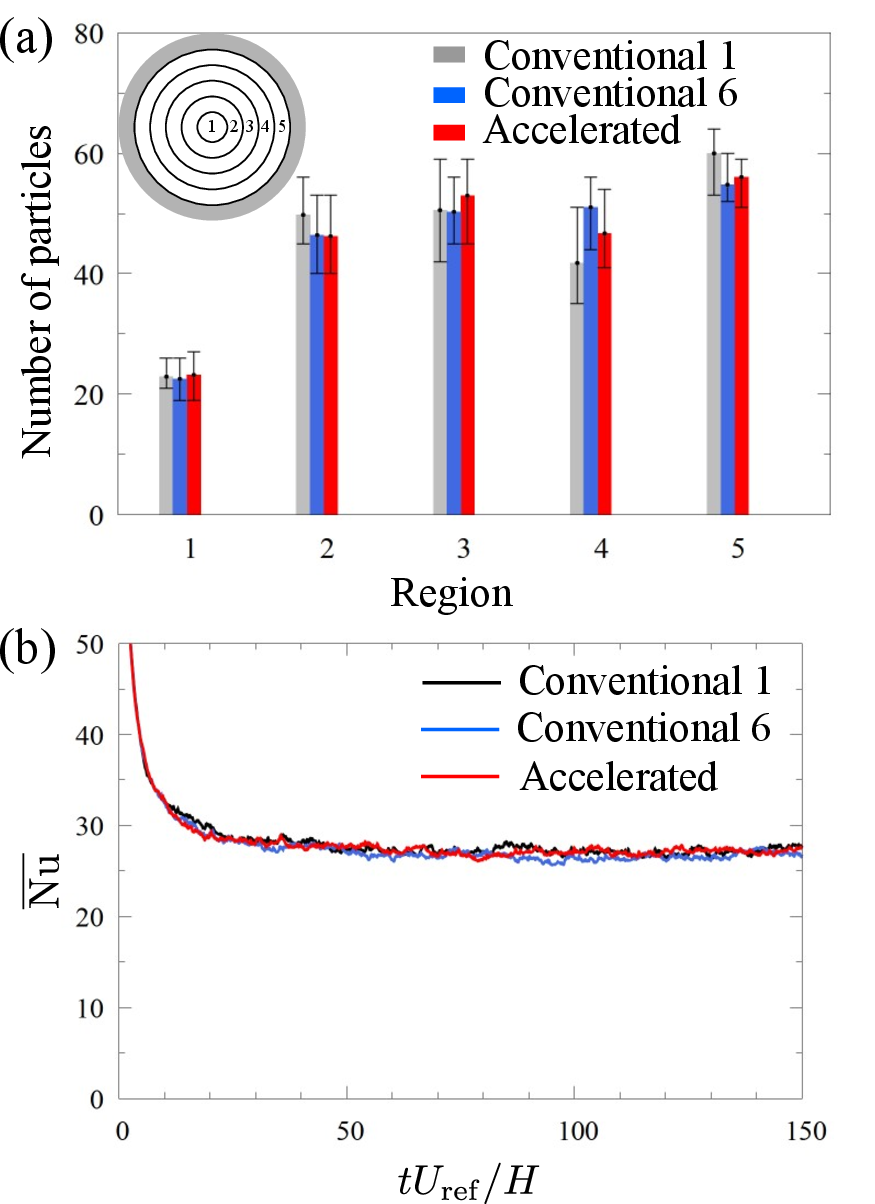} 
\caption{(a) Time-averaged distributions of particles, along with minimum/maximum value ranges during $140 \le tU_{\rm ref}/H \le 150$, 
evaluated within the divided regions on the cross-section of the circular tube; 
(b) time variation of the Nusselt number $\overline{\rm Nu}$ averaged over the circular tube wall.}
\label{fig:iceslurry_result2}
\end{figure}

In order to compare the results for Conventional 1 and 6 and Accelerated more quantitatively, we evaluate the time-averaged distributions of particles as shown in Fig.~\ref{fig:iceslurry_result2}(a). 
In this figure, the number of particles is counted within the divided regions of the circular tube, 
and each region (labeled as Region \#$m$) is defined as $0.1 H (m - 1) \le r \le 0.1 H m$ (where $r$ is the distance from the centerline of the circular tube). 
This figure indicates that the particle distributions for Conventional 1 and 6 and Accelerated are very close to each other. 
In addition, we evaluate the thermal boundary layer thickness by the Nusselt number, which is the non-dimensional value of the temperature gradient normal to the tube wall.
As the thermal boundary layer thickness decreases, the temperature varies near the tube wall more steeply, resulting in the increase of the Nusselt number. 
Fig.~\ref{fig:iceslurry_result2}(b) shows the time variation of the Nusselt number $\overline{\rm Nu}$ averaged over the tube wall. 
From this figure, we can see that the values of $\overline{\rm Nu}$ for Conventional 1 and 6 and Accelerated are nearly identical with each other, 
indicating that the thermal boundary layer thickness is also nearly identical. 
Thus, Conventional 1 and 6 and Accelerated give almost the same statistical results, even in a complex heat-transfer problem with hundreds of particles.

\begin{table}[t!]
\begin{center}
\caption{Mean and maximum {boundary-velocity and temperature errors of} the no-slip and isothermal conditions, 
computational time per time step, 
and proportion of the LBMs for flow and temperature fields, IBMs for flow and temperature fields, body motion, and other parts to the total computing time 
in the simulations of the ice slurry flow for the conventional and accelerated multi-direct-forcing IBMs.}
\label{tab:comp_iceslurry}
\scalebox{0.8}[0.8]{
  \begin{tabular}{llccc} \hline
                  &  & Conventional 1 & Conventional 6 & Accelerated  \\ 
                  &  & ($\omega = 1.0$, $\ell = 1$)  & ($\omega = 1.0$, $\ell = 6$) & ($\omega = C_{4}^{-1}$, $\ell = 1$)  \\ \hline
  	{Boundary-velocity error} & Mean & $1.232 \times 10^{-3}$ & $2.024 \times 10^{-4}$ & $2.819 \times 10^{-4}$ \\
	              & Maximum & $6.792 \times 10^{-2}$ & $1.655 \times 10^{-2}$ & $2.179 \times 10^{-2}$ \\ 
  	{Boundary-temperature error} & Mean & $1.178 \times 10^{-4}$ & $5.113 \times 10^{-5}$ & $6.310 \times 10^{-5}$ \\
	              & Maximum & $9.694 \times 10^{-3}$ & $6.085 \times 10^{-3}$ & $4.609 \times 10^{-3}$ \\ \hline
	Computational time (s) & per time step & 1.317 & 3.321 & 1.316  \\
	Proportion (\%) of	& LBM (flow) & \phantom{0}6.10  & \phantom{0}2.52 & \phantom{0}6.09 \\ 
						& LBM (temperature) & 10.60 & \phantom{0}4.29 & 10.60 \\ 
					& IBM (flow) & 41.34 & 60.32 & 41.40 \\
					& IBM (temperature) & 15.12 & 22.05 & 15.09 \\
					& Body motion & 26.80 & 10.81 & 26.78 \\
					& Other & \phantom{0}0.04 & \phantom{0}0.01 & \phantom{0}0.04 \\ \hline
  \end{tabular}}
 \end{center}
\end{table}

Table~\ref{tab:comp_iceslurry} shows the {velocity and temperature errors of} the no-slip and isothermal conditions and the computational time. 
It is noted that in this simulation, the isothermal condition on particle surfaces is enforced by the multi-direct-forcing IBM for the temperature field, and it is accelerated in the same way as that for the flow field. 
Thus, this table includes the errors in not only the no-slip condition but also the isothermal condition. 
The values of {boundary-velocity and temperature errors} are their time-averaged values in $9 \le tU_{\rm ref}/H \le 10$, where we cut off the initial transient data. 
The computational time is measured in $0 \le tU_{\rm ref}/H \le 10$.
These simulations were performed on 24 cores of a Xeon Silver 4214 (2.2 GHz) using MPI.
From this table, we can see that {boundary-velocity and temperature errors} are comparable between Conventional 6 and Accelerated, 
while the errors for Conventional 1 are much larger than the errors for Conventional 6 and Accelerated, 
which can be expected from the results shown in previous sections (e.g., see Fig.~\ref{fig:moving_circ_itr}(a)). 
The computational time for Accelerated is significantly reduced from that for Conventional 6 due to the reduction of the proportion of the IBMs for flow and temperature fields. 
Therefore, Accelerated produces results comparable to Conventional 6 while requiring significantly less computational cost.
The {large velocity and temperature errors of} the no-slip and isothermal conditions in Conventional 1 does not significantly affect the statistical results shown in Fig.~\ref{fig:iceslurry_result2}, since the interactions among hundreds of particles hide the influence of the {error}. 

Regarding the numerical stability, the parameter $A$ is defined by Eq.~(\ref{eq:A_sphere}), and $A = 1.07$ in this numerically stable simulation. 
When we set $\gamma = 0.9$, $A$ increases to $1.19$, resulting in numerical instability. 
Thus, there is an upper limit of $A$ for stable results, and its limit value is $A \lessapprox 1.0$, which is comparable to the results for a moving sphere in section~\ref{sec:sphere_Poiseuille}. 

In summary, even in a complex heat-transfer problem with many particles, 
the accelerated multi-direct-forcing IBMs for flow and temperature fields can give almost the same results as the conventional multi-direct-forcing IBMs while significantly reducing the computational cost.
Furthermore, the numerical stability of the body motion is determined by the value of $A$, and its limit value is $A \lessapprox 1.0$.

\section{Conclusions} \label{sec:conclusion}

This study investigated the {boundary-velocity error} and stability of the accelerated multi-direct-forcing immersed boundary method, which corresponds to the Richardson iteration for solving the matrix equation determining the volume force for rigid-body problems.
The optimal acceleration parameter $\omega$ for this method should be related to the characteristics of the matrix $\mathcal{A}$ in the matrix equation, which is determined by the weighting function for the interpolation and spreading procedures.
With Peskin's discrete delta function $\phi_s$ (whose support is $s\Delta{x}$ where $\Delta x$ is the grid spacing and $s=4$ or $s=3$), the infinity norm $\left\|\mathcal{A}\right\|_\infty$ and the constant $C_s$ ($C_4 = 0.375$ for $\phi_4$ and $C_3 = 0.50$ for $\phi_3$) have been previously proposed as candidates for determining the optimal acceleration parameter.

Our numerical results indicate that these values are very close to each other and the corresponding algorithms result in the same {velocity error of} the no-slip condition.
Our analysis of the discretized equations of body motion in moving boundary problems suggests that a single lumped parameter --- combining the acceleration parameter, the solid--fluid density ratio, and the spatial resolution of the immersed body --- should determine the numerical stability of the body motion.

First, we numerically examined the {velocity error of} the no-slip condition through stationary boundary problems, including a circular cylinder, an elliptical cylinder, and a sphere fixed in a planar Poiseuille flow.
As a result, setting the acceleration parameter as $\omega = C_{s}^{-1}$ is almost the optimal choice minimizing the {boundary-velocity error}, and the error for this choice is about ten times smaller than that for $\omega = 1$ without iteration.
Notably, this scenario holds independently of the distance between the neighboring boundary points, the shape of the boundary, the spatial dimensionality, and whether the weighting function is $\phi_4$ or $\phi_3$.

Second, we examined the numerical stability of simulations of moving boundary problems by considering a circular cylinder and a sphere moving in a planar Poiseuille flow and the sedimentation of a circular cylinder.
As a result, there is an upper limit of the lumped parameter for stable simulations, and this limit value is almost independent of the individual values of the acceleration parameter, the density ratio, the spatial resolution, and the gravitational acceleration.
Additionally, we showed how the numerical stability is affected by the number of iterations of the multi-direct-forcing method.

Finally, we applied the accelerated multi-direct-forcing immersed boundary method to more complex problems: butterfly flight and ice slurry flow. 
As a result, the accelerated multi-direct-forcing method can give almost the same results as the conventional multi-direct-forcing method while significantly reducing the computational cost.
Furthermore, the numerical stability of the body motion is determined by the lumped parameter in the same way as seen for a moving circular cylinder or a moving sphere. 

Our work provides quantitative guidelines for the \textit{a priori} choice of simulation parameters leading to both numerically stable and accurate simulations of moving boundary problems using the accelerated multi-direct-forcing immersed boundary method.


\section*{Acknowledgment}
{KS was supported by JSPS KAKENHI grant number JP23H01341.}
TK received funding from the European Research Council (ERC) under the European Union's Horizon 2020 research and innovation program (803553).
For the purpose of open access, the authors have applied a Creative Commons Attribution (CC BY) license to any Author Accepted Manuscript version arising from this submission.
KS thanks Mr.~S.~Tsujimura for providing the data for the ice slurry flow in Figs.~\ref{fig:iceslurry_result1} and \ref{fig:iceslurry_result2}. 

\appendix

\section{Implementation with the lattice Boltzmann method} \label{sec:algorithm}

In this section, we describe the full algorithm for moving boundary problems implemented with the lattice Boltzmann method (LBM) based on Ref.~\cite{Suzuki2011, Inamuro2021}.
We use the D2Q9 lattice for two-dimensional (2D) simulations and the D3Q15 lattice for three-dimensional (3D) simulations.
In the following, we describe the algorithm for a 2D rigid body moving in a 2D flow field, and we use the non-dimensional variables defined in \ref{sec:nondim}.
Note that the same notation as in section~\ref{sec:method} is used for the non-dimensional variables.
We can simply extend the 2D algorithm to the 3D one, except for the description of the body motion. 
If the body is a 3D rigid body, we must apply a 3D coordinate transformation using Euler angles or quaternions. 
The description is complicated so is omitted here. 
For more details about the equations of motion of a 3D rigid body, please see Ref.~\cite{Suzuki2011}.

We use a weak coupling scheme which alternately solves for the body and fluid dynamics, both with the same time step $\Delta{t}$.
In addition, we use the diffusive time scale for the LBM, i.e., $\Delta{t} = {\rm Sh} \Delta{x}$, where ${\rm Sh}$ is the Strouhal number and $\Delta{x}$ is the lattice spacing.
Thus, even in the equation of body motion, the time derivative terms are accompanied by ${\rm Sh}$.
See also \ref{sec:nondim}.

In the present study, we use the forcing scheme which is consistent with the fractional-step approach given in Eq.~(\ref{eq:uast}) and (\ref{eq:frac2}).
Note that the forcing scheme proposed by Guo et al.~\cite{Guo2002} is not exactly consistent with Eq.~(\ref{eq:uast}) and (\ref{eq:frac2}) because it includes two forcing steps where not only the distribution functions but also the flow velocity itself are corrected by the volume force~\cite{Kang2011}.

\subsection{Full algorithm for moving boundary problems} \label{sec:fullalgorithm}

\begin{enumerate}
 \item Starting from the data at time $t$, update the velocity $\bm{U}_{\rm c}$ and angular velocity ${\Omega}_{\rm c}$ of the center of mass (COM) of the body by
 \begin{align}
  M {\rm Sh}\frac{\bm{U}_{\rm c}(t+\Delta{t}) - \bm{U}_{\rm c}(t)}{\Delta{t}} &=  \bm{F}_{\rm tot}(t) + \bm{F}_{\rm in}(t) + M \left( 1- \frac{1}{\gamma} \right)\bm{G}, \label{eq:Euler1}\\
  {I} {\rm Sh}\frac{{\Omega}_{\rm c}(t+\Delta{t}) - {\Omega}_{\rm c}(t)}{\Delta{t}} &= {T}_{\rm tot}(t) + {T}_{\rm in}(t),\label{eq:Euler2}
 \end{align}
 where $M$ and $I$ are the mass and inertial moment of the body, respectively, and $\bm{G}$ is the gravitational acceleration. 
 The fluid force $\bm{F}_{\rm tot} + \bm{F}_{\rm in}$ and torque ${T}_{\rm tot} + {T}_{\rm in}$ are given by
 \begin{align}
  \bm{F}_\text{tot}(t)&= -\sum_{\bm{x}}\bm{g}(\bm{x},t) (\Delta{x})^2, \label{eq:fex_sabun} \\
  {T}_\text{tot}(t)&= -\sum_{\bm{x}}\left[ (x-{X}_\text{c}(t)) {g}_y(\bm{x},t) - (y-{Y}_\text{c}(t)) {g}_x(\bm{x},t) \right] (\Delta{x})^2 \label{eq:tex_sabun} 
 \end{align}
 and 
 \begin{align}
  \bm{F}_\text{in}(t) &= {\rm Sh}\frac{\bm{P}_\text{in}(t)-\bm{P}_\text{in}(t-\Delta{t})}{\Delta t}, \\
  {T}_\text{in}(t) &= {\rm Sh}\frac{{L}_\text{in}(t)-{L}_\text{in}(t-\Delta{t})}{\Delta t}, \label{eq:fin_Lp}
 \end{align}
 where $\bm{P}_{\rm in}$ and $L_{\rm in}$ are the linear and angular momentum of the fluid inside the body.
 The Lagrangian point approximation is used for the internal mass effects $\bm{F}_{\rm in}$ and $T_{\rm in}$ (see Ref.~\cite{Suzuki2011} for more detail).

 \item Update the COM position $\bm{X}_{\rm c}$ and angle $\Theta_{\rm c}$ of the body by
 \begin{align}
  {\rm Sh} \frac{\bm{X}_{\rm c}(t+\Delta{t}) - \bm{X}_{\rm c}(t)}{\Delta t} &= \bm{U}_\text{c}(t), \label{eq:Euler3} \\
  {\rm Sh} \frac{\Theta_{\rm c}(t+\Delta{t}) - \Theta_{\rm c}(t)}{\Delta t} &= \Omega_{\rm c}(t). \label{eq:Euler4} 
 \end{align}

 \item Update the position $\bm{X}_k$ and velocity $\bm{U}_k$ of the boundary points by
 \begin{align}
  \bm{X}_{k}(t+{\Delta t}) &= \bm{X}_{\rm c}(t+\Delta t) + \mathcal{R}(t+ \Delta t) \cdot \left[ \bm{X}_{k}(0)-\bm{X}_{\rm c}(0) \right], \label{eq:update_Xk} \\
  \bm{U}_{k}(t+{\Delta t}) &= \bm{U}_{\rm c}(t+\Delta t) + {\rm Sh}\frac{\text{d} \mathcal{R}}{\text{d}t}(t+ \Delta t) \cdot\left[ \bm{X}_{k}(0)-\bm{X}_{\rm c}(0) \right], \label{eq:update_Uk}
 \end{align}
 where $\mathcal{R}$ is the two-dimensional rotational matrix given by
 \begin{align}
  \mathcal{R}(t) = \left[
  \begin{array}{rr}
   \cos\Theta_{\rm c}(t) & -\sin\Theta_{\rm c}(t)  \\
   \sin\Theta_{\rm c}(t) &  \cos\Theta_{\rm c}(t)  
  \end{array}
  \right].
 \end{align}
 The time derivative $\text{d} \mathcal{R}/\text{d}t$ of this matrix is given by
 \begin{align}
  {\rm Sh}\frac{\text{d} \mathcal{R}}{\text{d}t}(t) = \Omega_{\rm c}(t) \left[
  \begin{array}{rr}
   - \sin\Theta_{\rm c}(t) & -\cos\Theta_{\rm c}(t)  \\
   \cos\Theta_{\rm c}(t) &  - \sin\Theta_{\rm c}(t)  
  \end{array}
  \right].
 \end{align}

 \item Compute the temporary distribution functions $f^*_i$ ($i=1,\ldots, 9$) by
 \begin{equation}
  f^*_i(\bm{x}+\bm{c}_i \Delta{x},t+\Delta{t}) = f_i(\bm{x},t) - \frac{1}{\tau}\left[ f_i(\bm{x},t) - f^{\rm eq}_i(p(\bm{x},t),\bm{u}(\bm{x},t))\right], 
 \end{equation}
 where $\bm{c}_i$ is the lattice velocity, 
 and $f^{\rm eq}_i$ is the local equilibrium distribution function given as
 \begin{equation}
  f^{\rm eq}_i (p,\bm{u}) = E_i \left[ 3p + 3\bm{c}_i \cdot \bm{u} + \frac{9}{2}(\bm{c}_i \cdot \bm{u})^2 - \frac{3}{2}\bm{u} \cdot \bm{u}\right],
 \end{equation}
 where $p$ and $\bm{u}$ are the pressure and flow velocity, respectively, and $E_i$ is the weight coefficient depending on the lattice model.

 \item Compute the temporary velocity $\bm{u}^*$ by
 \begin{equation}
  \bm{u}^*(\bm{x},t+\Delta{t}) = \sum_{i=1}^9 \bm{c}_i f^*_i(\bm{x},t+\Delta{t}).
 \end{equation}

 \item Compute the volume force $\bm{g}$ through the accelerated multi-direct-forcing IBM as follows:
 \begin{description}
  \setlength{\leftskip}{10mm}
  \item[Step 0.]
  Compute the initial volume force at the boundary points of the body: 
  \begin{equation}
   \bm{g}_1(\bm{X}_k,t+\Delta{t})= \omega \rho_{\rm f} {\rm Sh}\frac{\bm{U}_k-\bm{u}^*(\bm{X}_k,t+\Delta{t})}{\Delta{t}}. \label{eq:step0}
  \end{equation}
  \item[Step 1.]
  Compute the volume force at the lattice points during the $\ell $th iteration:
  \begin{equation}
   \bm{g}_{\ell }(\bm{x},t+\Delta{t})=\sum_{k=1}^{N}\bm{g}_{\ell }(\bm{X}_k,t+\Delta{t})\ W(\bm{x}-\bm{X}_k)\ \Delta{V}_k. \label{eq:distri}
  \end{equation}
  \item[Step 2.]
  Correct the flow velocity at the lattice points:
  \begin{equation}
   \bm{u}_{\ell}(\bm{x},t+\Delta{t})=\bm{u}^*(\bm{x},t+\Delta{t})+ \frac{1}{\rho_{\rm f}} \frac{\Delta{t}}{{\rm Sh}}\bm{g}_{\ell }(\bm{x},t+\Delta{t}).
  \end{equation}
  \item[Step 3.]
  Interpolate the flow velocity at the boundary {points} of the body:
  \begin{equation}
   \bm{u}_{\ell}(\bm{X}_k,t+\Delta{t})=\sum_{\bm{x}}\bm{u}_{\ell}(\bm{x},t+\Delta{t})\ W(\bm{x}-\bm{X}_k)\ (\Delta{x})^2.
  \end{equation}
  \item[Step 4.]
  If the error in $\bm{u}_{\ell}(\bm{X}_k,t+\Delta{t})$ from the no-slip condition is not sufficiently small, 
  then correct the volume force at the boundary points of the body and return to {\bf Step 1}:
  \begin{equation}
   \bm{g}_{\ell +1}(\bm{X}_k,t+\Delta{t})=\bm{g}_{\ell }(\bm{X}_k,t+\Delta{t})+ \omega \rho_{\rm f} {\rm Sh}\frac{\bm{U}_k-\bm{u}_{\ell}(\bm{X}_k,t+\Delta{t})}{\Delta{t}}. \label{eq:step4}
  \end{equation}
  If the error of the no-slip boundary condition is sufficiently small, then go to {\bf{Step 5}}. 
  \item[Step 5.]
  Determine the volume force at the lattice points at time $(t+\Delta{t})$:
  \begin{equation}
   \bm{g}(\bm{x},t+\Delta{t})=\bm{g}_{\ell}(\bm{x},t+\Delta{t}). \label{eq:step5}
  \end{equation}
 \end{description}

 \item Update the distribution functions $f_i$ by
 \begin{equation}
  f_i(\bm{x},t+\Delta{t})=f^*_i(\bm{x},t+\Delta{t}) + 3\Delta{x}E_i\bm{c}_i \cdot \bm{g}(\bm{x},t+\Delta{t}). \label{eq:kousin}
 \end{equation}

 \item Update the pressure $p$ and flow velocity $\bm{u}$ by
 \begin{align}
  p(\bm{x},t+\Delta{t}) &= \frac{1}{3}\sum_{i=1}^{9} f_i(\bm{x},t+\Delta{t}), \\
  \bm{u}(\bm{x},t+\Delta{t}) &= \sum_{i=1}^9 \bm{c}_i f_i(\bm{x},t+\Delta{t}).
 \end{align}
\end{enumerate}

\subsection{Non-dimensional variables} \label{sec:nondim}

In \ref{sec:fullalgorithm}, we use the following non-dimensional variables defined by a characteristic length $\hat{L}$, 
a characteristic particle speed $\hat{c}$, a characteristic time scale $\hat{t}_0 = \hat{L}/\hat{U}$ where $\hat{U}$ is a characteristic flow speed, 
and a fluid density $\hat{\rho}_\text{f}$: 
\begin{align}
 \left. \begin{array}{lll}
  \bm{c}_i = \hat{\bm{c}}_i/\hat{c}, & \bm{x} = \hat{\bm{x}}/\hat{L}, & t=\hat{t}/\hat{t}_0, \\
  \Delta{x} = \Delta{\hat{x}}/\hat{L}, & \Delta{t} = \Delta{\hat{t}}/\hat{t}_0, &\\
  f_i=\hat{f}_i/\hat{\rho}_\text{f}, & \bm{u} = \hat{\bm{u}}/\hat{c}, & p = \hat{p}/(\hat{\rho}_\text{f}\hat{c}^2), \\
  \nu = \hat{\nu }/(\hat{c}\hat{L}), & \bm{g} = \hat{\bm{g}}\hat{L}/(\hat{\rho}_\text{f}\hat{c}^2), &  \\
  \bm{X}_k= \hat{\bm{X}}_k/\hat{L} & \bm{U}_k= \hat{\bm{U}}_k/\hat{c}, & \\ 
  \bm{X}_\text{c}= \hat{\bm{X}}_\text{c}/\hat{L} & \bm{U}_\text{c}= \hat{\bm{U}}_\text{c}/\hat{c}, & {\Omega}_\text{c} = \hat{{\Omega}}_\text{c}\hat{L}/\hat{c},  \\ 
  M=\hat{M}/(\hat{\rho}_\text{f}\hat{L}^d), & {I}=\hat{{I}}/(\hat{\rho}_\text{f} \hat{L}^{d+2}), &  \\
  \bm{F} = \hat{\bm{F}}/(\hat{\rho}_\text{f}\hat{c}^2\hat{L}^{d-1}), & {T} = \hat{{T}}/(\hat{\rho}_\text{f}\hat{c}^2\hat{L}^{d}), & \\
  \bm{F}_\text{tot} = \hat{\bm{F}}_\text{tot}/(\hat{\rho}_\text{f}\hat{c}^2\hat{L}^{d-1}), & {T}_\text{tot} = \hat{{T}}_\text{tot}/(\hat{\rho}_\text{f}\hat{c}^2\hat{L}^{d}), & \\
  \bm{F}_\text{in} = \hat{\bm{F}}_\text{in}/(\hat{\rho}_\text{f}\hat{c}^2\hat{L}^{d-1}), & {T}_\text{in} = \hat{{T}}_\text{in}/(\hat{\rho}_\text{f}\hat{c}^2\hat{L}^{d}), & \\
  \bm{P}_\text{in} = \hat{\bm{P}}_\text{in}/(\hat{\rho}_\text{f}\hat{c}\hat{L}^{d}), & {L}_\text{in} = {{L}}_\text{in}/(\hat{\rho}_\text{f}\hat{c}\hat{L}^{d+1}), &
 \end{array}
 \right\} \label{eq:dim1}
\end{align}
where the circumflex represents `dimensional,' and $d$ represents the spatial dimension ($d=2$ and $3$ in the two- and three-dimensional cases, respectively).

In the LBM, the time step $\Delta{\hat{t}}$ is equal to the time span during which the fluid particles travel one lattice spacing, 
that is, $\Delta{\hat{x}}/\Delta{\hat{t}}=\hat{c}$.
The non-dimensional lattice spacing is $\Delta{x} = \Delta{\hat{x}}/\hat{L}$,
and the non-dimensional time step is $\Delta{t}=\Delta{\hat{t}}/(\hat{L}/\hat{c})$ in the diffusive time scale. 
Thus, we have $\Delta t = {\rm Sh} \Delta x$ (where ${\rm Sh} = \hat{L}/(\hat{t}_0 \hat{c}) = \hat{U}/\hat{c}$ is the Strouhal number) in the diffusive time scale.

\section{Volume force with iteration} \label{sec:iteration}

When we iterate the procedure of the accelerated multi-direct-forcing IBM, the volume force is given by the following recurrence relation:
\begin{equation}
\bm{g}^n_{\ell +1}(\bm{X}_k) = \bm{g}^{n}_{\ell}(\bm{X}_k) + \omega \rho_{\rm f} \frac{\bm{U}_k^n - \bm{u}_{\ell} (\bm{X}_k)}{{\Delta{t}}}, \label{eq:giterate}
\end{equation}
where 
\begin{align}
&\bm{g}^{n}_{0}(\bm{X}_k) = \bm{0}, \\
&\bm{u}_0(\bm{X}_k) = \bm{u}^*(\bm{X}_k), \\
&\bm{u}_{\ell }(\bm{X}_k) = \bm{u}^*(\bm{X}_k) + \frac{1}{\rho_{\rm f}}\tilde{\bm{g}}^n_{\ell}(\bm{X}_k) {\Delta{t}} \label{eq:ul}, 
\end{align}
and $\tilde{\bm{g}}^n_{\ell}(\bm{X}_k)$ is given by
\begin{align}
&\tilde{\bm{g}}^n_{\ell}(\bm{X}_k) = \sum_{\bm{x}}{\bm{g}}^n_{\ell}(\bm{x})W(\bm{x}-\bm{X}_k) (\Delta{x})^d, \label{eq:gtilde}\\
&{\bm{g}}^n_{\ell}(\bm{x}) = \sum_{k=1}^N {\bm{g}}^n_{\ell}(\bm{X}_k)W(\bm{x}-\bm{X}_k) \Delta{V}_k. \label{eq:gx}
\end{align}
Note that $\tilde{\bm{g}}^n_{\ell}(\bm{X}_k)$ is not generally equal to ${\bm{g}}^n_{\ell}(\bm{X}_k)$
duo to the interpolation and spreading procedures.

From Eqs.~(\ref{eq:giterate})--(\ref{eq:ul}), we have the recurrence relation of $\bm{F}^n$ as follows:
\begin{align}
&\underbrace{\sum_{k=1}^N \bm{g}^n_{\ell +1}(\bm{X}_k)\Delta{V}_k}_{- \bm{F}^n_{\ell +1}} 
= \underbrace{\sum_{k=1}^N\bm{g}^{n}_{\ell}(\bm{X}_k)\Delta{V}_k}_{-\bm{F}^n_{\ell}} + \omega \rho_{\rm f} \sum_{k=1}^N \frac{\bm{U}_k^n - \bm{u}_{\ell} (\bm{X}_k)}{{\Delta{t}}} \Delta{V}_k, \notag \\
\Longrightarrow~~&\bm{F}_{\ell+1}^n  = \bm{F}^n_{\ell}  - \underbrace{ \omega \rho_{\rm f} \sum_{k=1}^N  \frac{\bm{U}_k^n - \bm{u}^{*} (\bm{X}_k)}{{\Delta{t}}} \Delta{V}_k}_{- \bm{F}_1^n} + \omega \underbrace{\sum_{k=1}^N  \tilde{\bm{g}}^n_{\ell}(\bm{X}_k) \Delta{V}_k}_{- \tilde{\bm{F}}^n_\ell}, \notag \\
\Longrightarrow~~&\bm{F}_{\ell+1}^n  = \bm{F}^n_{\ell} + \bm{F}_1^n - \omega \tilde{\bm{F}}^n_\ell. \label{eq:fell}
\end{align}
Note that $\tilde{\bm{F}}^n_\ell$ is not generally equal to $\bm{F}^n_{\ell}$.

Now consider the relationship between $\tilde{\bm{F}}^n_\ell$ and $\bm{F}^n_{\ell}$.
By using Eq.~(\ref{eq:gtilde}) and (\ref{eq:gx}), we develop
\begin{align}
 \tilde{\bm{F}}^n_\ell &= - \sum_{k=1}^N\tilde{\bm{g}}^{n}_{\ell}(\bm{X}_k)\Delta{V}_k, \notag \\
 &=- \sum_{k=1}^N \left[ \sum_{\bm{x}}{\bm{g}}^n_{\ell}(\bm{x})W(\bm{x}-\bm{X}_k) (\Delta{x})^d \right]\Delta{V}_k, \notag \\
 &=- \sum_{k=1}^N  \sum_{\bm{x}} \left[ \sum_{j=1}^N {\bm{g}}^n_{\ell}(\bm{X}_j)W(\bm{x}-\bm{X}_j) \Delta{V}_j \right] W(\bm{x}-\bm{X}_k) (\Delta{x})^d \Delta{V}_k, \notag \\
 &=- \sum_{k=1}^N \sum_{j=1}^N {\bm{g}}^n_{\ell}(\bm{X}_j) \Delta{V}_j \underbrace{\sum_{\bm{x}}W(\bm{x}-\bm{X}_j) W(\bm{x}-\bm{X}_k) (\Delta{x})^d \Delta{V}_k}_{A_{jk}}, \notag \\
 &= - \sum_{j=1}^N {\bm{g}}^n_{\ell}(\bm{X}_j) \Delta{V}_j \underbrace{\sum_{k=1}^N A_{jk}}_{a_j}, \notag \\
 &= - \sum_{j=1}^N a_j {\bm{g}}^n_{\ell}(\bm{X}_j) \Delta{V}_j. \label{eq:mid}
\end{align}
As shown in section~\ref{sec:matrix}, the coefficient $a_j$ can be well approximated by $\lambda_{\max}$.
Thus, Eq.~(\ref{eq:mid}) can be approximated by
\begin{align}
 \tilde{\bm{F}}^n_\ell \simeq - \sum_{j=1}^N \lambda_{\rm max} {\bm{g}}^n_{\ell}(\bm{X}_j) \Delta{V}_j = \lambda_{\max} \bm{F}^n_{\ell}. \label{eq:approx}
\end{align}

By substituting Eq.~(\ref{eq:approx}) to Eq.~(\ref{eq:fell}), we have
\begin{align}
 \bm{F}_{\ell+1}^n  \simeq  (1- \lambda_{\max} \omega ) \bm{F}^n_{\ell} + \bm{F}_1^n, 
\end{align}
This simple recurrence relation can be readily solved, and the solution is 
\begin{align}
 \bm{F}_{\ell}^n \simeq {\frac{1-(1-\lambda_{\max} \omega )^{\ell+1}}{\lambda_{\max} \omega }} \bm{F}_1^n. \label{eq:eta}
\end{align}
Thus, the effect of the iteration is confined to the coefficient of $\bm{F}_1^n$ in Eq.~(\ref{eq:eta}) given by
\begin{equation}
 \eta = \frac{1-(1-\lambda_{\max} \omega )^{\ell+1}}{\lambda_{\max} \omega }.
\end{equation}
Fig.~\ref{fig:eta} shows $\eta$ as a function of $\lambda_{\rm max} \omega$ for various iteration counts $\ell$.
Evidently, $\eta$ is constant at $1$ for $\ell=1$ and monotonically decreases with increasing $\lambda_{\rm max} \omega$ for $\ell >1$, 
and $\eta=1$ when $\lambda_{\rm max} \omega = 1$ for any value of $\ell$.
Thus, $\eta \ge 1$ for $\lambda_{\rm max} \omega \le 1$.

\setcounter{figure}{0}
\begin{figure}[!tb]
 \centering
 \includegraphics[width=8cm,clip]{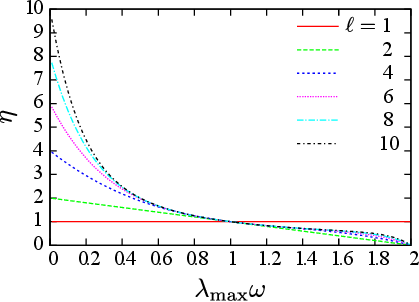} 
 \caption{Relationship between $\eta$ and $\lambda_{\rm max} \omega$ for various iteration counts $\ell$.}
 \label{fig:eta}
\end{figure}

Therefore, the force acting on the body when the iteration count is $\ell$ is estimated by
\begin{equation}
\bm{F}_{\ell}^n \simeq \eta \bm{F}_1^n = - \eta \omega \rho_{\rm f} \frac{S \Delta{x}}{N}\sum_{k=1}^{N}\frac{\bm{U}_k^n - \bm{u}^*(\bm{X}_k)}{{\Delta{t}}}. \label{eq:gsum2_2}
\end{equation}


\begin{thebibliography}{00}


\bibitem{Peskin1972}
  C. S. Peskin,
  Flow patterns around heart valves: A numerical method,
  \textit{J. Comput. Phys.} 10 (1972) 252--271.

\bibitem{Peskin1977}
  C. S. Peskin,
  Numerical analysis of blood flow in the heart,
  \textit{J. Comput. Phys.} 25 (1977) 220--252.

\bibitem{Mittal2005}
  R. Mittal, G. Iaccarino,
  Immersed boundary methods,
  \textit{Annu. Rev. Fluid Mech.} 37 (2005) 239--261.

\bibitem{Griffith2020}
  B. E. Griffith, N. A. Patankar,
  Immersed methods for fluid--structure interaction,
  \textit{Annu. Rev. Fluid Mech.} 52 (2020) 421--428.

 \bibitem{Verzicco2023}
  R. Verzicco,
  Immersed boundary methods: historical perspective and future outlook,
  \textit{Annu. Rev. Fluid Mech.} 55 (2023) 129--155. 

\bibitem{Mohd1997}
  J. Mohd-Yusof,
  Combined immersed-boundary/B-spline methods for simulations of flow in complex geometries,
  \textit{CTR Annual Research Briefs} (1997) 317--327.

\bibitem{Ye1999}
T. Ye, R. Mittal, H.S. Udaykumar, W. Shyy,
An accurate Cartesian grid method for viscous incompressible flows with complex immersed boundaries,
\textit{J. Comput. Phys.} 156 (1999) 209--240.

\bibitem{Mittal2008}
R. Mittal, H. Dong, M. Bozkurttas, F.M. Najjar, A. Vargas, A. von Loebbecke,
A versatile sharp interface immersed boundary method for incompressible flows with complex boundaries,
\textit{J. Comput. Phys.} 227 (2008) 4825--4852.


\bibitem{Peskin2002}
  C. S. Peskin, 
  The immersed boundary method, 
  \textit{Acta Numer.} 11 (2002) 479--517. 

\bibitem{Uhlmann2005}
  M. Uhlmann,
  An immersed boundary method with direct forcing for the simulation of particulate flows,
  \textit{J. Comput. Phys.} 209 (2005) 448--476.

 \bibitem{Wang2008}
	Z. Wang, J. Fan, K. Luo,
	Combined multi-direct forcing and immersed boundary method for simulating flows with moving particles,
	\textit{Int. J. Multiphase Flow} 34	(2008) 283--302.

\bibitem{Wu2009}
  J. Wu, C. Shu, 
  Implicit velocity correction-based immersed boundary--lattice Boltzmann method and its applications,
  \textit{J. Comput. Phys.} 228 (2009) 1963--1979.

\bibitem{Hu2014}
  Y. Hu, H. Yuan, S. Shu, X. Niu, M. Li, 
  An improved momentum exchanged-based immersed boundary--lattice Boltzmann method by using an iterative technique,
  \textit{Comput. Math. Appl.} 68 (3) (2014) 140--155.

\bibitem{Zhang2020}
  Y. Zhang, G. Pan, Y. Zhang, S. Haeri, 
  A relaxed multi-direct-forcing immersed boundary-cascaded lattice Boltzmann method accelerated on GPU,
  \textit{Comput. Phys. Commun.} 248 (2020) 106980.

 \bibitem{Gsell2019}
	S. Gsell,  U. D'Ortona, J. Favier,
	Explicit and viscosity-independent immersed-boundary scheme for the lattice Boltzmann method,
	\textit{Phys. Rev. E} 100 (2019) 033306.

 \bibitem{Gsell2021}
	S. Gsell, J. Favier,
	Direct-forcing immersed-boundary method: A simple correction preventing boundary slip error,
	\textit{J. Comput. Phys.} 435 (2021) 110265.

\bibitem{Kruger2017}
T. Kr\"{u}ger, H. Kusumaatmaja, A. Kuzmin, O. Shardt, G. Silva, E. M. Viggen, 
\textit{The Lattice Boltzmann Method: Principles and Practice},
Springer, Berlin, 2017.

\bibitem{Inamuro2021}
  T. Inamuro, M. Yoshino, K. Suzuki, 
  \textit{An Introduction to the Lattice Boltzmann Method: A Numerical Method for Complex Boundary and Moving Boundary Flows}, 
  World Scientific Publishing, 2021.


 \bibitem{Goldstein1993}
 D. Goldstein, R. Handler, L. Sirovich, 
 Modeling a no-slip flow boundary with external force field, 
 {\it J. Comput. Phys.} 105 (1993) 354--66.


\bibitem{Suzuki2011}
  K. Suzuki, T. Inamuro,
  Effect of internal mass in the simulation of a moving body by the immersed boundary method, 
  \textit{Comput. Fluids} 49 (2011) 173--187.


 \bibitem{Feng2009}
	Z.-G. Feng, E. E. Michaelides,
	Robust treatment of no-slip boundary condition and velocity updating for the lattice--Boltzmann simulation of particulate flows,
	\textit{Comput. Fluids} 38 (2009) 370--381.

 

 \bibitem{Borazjani2008}
 I. Borazjani, L. Ge, F. Sotiropoulos, 
 Curvilinear immersed boundary method for simulating fluid structure interaction with complex 3D rigid bodies, 
 {\it J. Comput. Phys.} 227 (2008) 7587--7620.

 \bibitem{Lacis2016}
 U. L\={a}cis, K. Taira, S. Bagheri, 
 A stable fluid–structure-interaction solver for low-density rigid bodies using the immersed boundary projection method, 
 {\it J. Comput. Phys.} 305 (2016) 300--318.
 

 

\bibitem{Suzuki2024a}
  K. Suzuki, D. Iguchi, K. Ishizaki, M. Yoshino,
  Bottom-up butterfly model with thorax-pitch control and wing-pitch flexibility,
  \textit{Bioinsp. Biomim.} 19 (2024) 046019.

\bibitem{Suzuki2024b}
  K. Suzuki, R. Uchida, R. Shiomi, T. Asaoka, M. Yoshino,
  Microconvection effects on the cooling performance of ice slurry flows within a circular tube: Immersed boundary--lattice Boltzmann simulations,
  \textit{Int. J. Heat Mass Transfer} 232 (2024) 125953.


\bibitem{Suzuki2021}
  K. Suzuki, T. Kuroiwa, T. Asaoka, M. Yoshino,
  Particle-resolved simulations of ice slurry flows in a square duct by the thermal immersed boundary--lattice Boltzmann method,
  \textit{Comput. Fluids} 228 (2021) 105064.

\bibitem{Guo2002}
Z. Guo, C. Zheng, B. Shi,
{Discrete lattice effects on the forcing term in the lattice Boltzmann method},
\textit{Phys. Rev. E} 65, (2002) 046308. 

\bibitem{Kang2011}
  S. Kang, Y. A. Hassan,
  A comparative study of direct-forcing immersed boundary--lattice Boltzmann methods for stationary complex boundaries,
  \textit{Int. J. Numer. Meth. Fluids} 66 (2011) 1132--1158.


\end{thebibliography}
\end{document}